\documentclass{aa}
\usepackage{natbib}
\bibpunct{(}{)}{;}{a}{}{,}
\usepackage{epstopdf}
\usepackage{epsfig}
\usepackage{subfig}
\usepackage{url}
\usepackage{rotating}
\usepackage{lscape}
\usepackage{lipsum}
\usepackage{supertabular}
\usepackage{longtable}
\usepackage{graphicx}
\usepackage{lscape}
\usepackage{longtable,lscape}

\usepackage{txfonts}
\usepackage{natbib,twoopt}
\bibpunct{(}{)}{;}{a}{}{,} 
\newcommandtwoopt{\citeads}[3][][]{\href{http://adsabs.harvard.edu/abs/#3}%
{\citealp[#1][#2]{#3}}}
\newcommandtwoopt{\citepads}[3][][]{\href{http://adsabs.harvard.edu/abs/#3}%
{\citep[#1][#2]{#3}}}
\newcommandtwoopt{\citetads}[3][][]{\href{http://adsabs.harvard.edu/abs/#3}%
{\citet[#1][#2]{#3}}}
\newcommandtwoopt{\citeyearads}[3][][]%
{\href{http://adsabs.harvard.edu/abs/#3}{\citeyear[#1][#2]{#3}}}

\begin{document}

\title{Near-Earth asteroids spectroscopic survey at Isaac Newton Telescope}
 
\author
{
 M. Popescu\inst{1,2,3}
 \and
 O. Vaduvescu\inst{4,1}
 \and
 J. de Le\'on\inst{1,2}
 \and
 R. M. Gherase\inst{3,5}
 \and
 J. Licandro\inst{1,2}
 \and
 I. L. Boac\u{a}\inst{3}
 \and
 A. B. \c{S}onka\inst{3,6}
 \and
 R. P. Ashley\inst{4}
 \and
 T. Mo{\v c}nik\inst{4,7}
 \and
 D. Morate\inst{8,1}
 \and
 M. Predatu\inst{9}
 \and
 M\'ario De Pr\'a\inst{10}
 \and
 C. Fari{\~n}a\inst{4,1}
 \and 
 H. Stoev\inst{4,11}
 \and
 M. D\'{\i}az Alfaro\inst{4,12}
 \and
 I. Ordonez-Etxeberria\inst{4,13}
 \and
  F. L{\'o}pez-Mart{\'{\i}}nez\inst{4}
 \and
 R. Errmann\inst{4}
 }
\offprints{M. Popescu, \email{mpopescu@iac.es}}
 \institute{ Instituto de Astrof\'{\i}sica de Canarias (IAC), C/V\'{\i}a L\'{a}ctea s/n, 38205 La Laguna, Tenerife, Spain
 \and
 Departamento de Astrof\'{\i}sica, Universidad de La Laguna, 38206 La Laguna, Tenerife, Spain
 \and
 Astronomical Institute of the Romanian Academy, 5 Cu\c{t}itul de Argint, 040557 Bucharest, Romania
 \and
 Isaac Newton Group of Telescopes (ING), Apto. 321, E-38700 Santa Cruz de la Palma, Canary Islands, Spain
 \and
  Astroclubul Bucure\c{s}ti, B-dul LascarCatargiu 21, sect 1, Bucharest, Romania
 \and
  Faculty of Physics, Bucharest University, 405 Atomistilor str, 077125 Magurele, Ilfov, Romania
 \and
  Department of Earth and Planetary Sciences, University of California, Riverside, CA 92521, USA  
 \and
  Observat\'orio Nacional, Coordenaç\~ao de Astronomia e Astrof\'{\i}sica, 20921-400 Rio de Janeiro, Brazil
 \and
 Faculty of Sciences, University of Craiova, Romania
 \and
 Florida Space Institute, University of Central Florida, Orlando, FL 32816, USA
 \and
  Fundaci\'{o}n Galileo Galilei -- INAF, Rambla Jos\'{e} Ana Fern\'{a}ndez P\'{e}rez, 7, 38712 Bre\~{n}a Baja, Spain
 \and 
  National  Solar  Observatory,  3665  Discovery  Drive,  Boulder,  CO 80303, USA
 \and   
 Dpto. de F\'isica Aplicada I, Universidad del Pa\'is Vasco / Euskal Herriko Unibertsitatea
}

\abstract
{
 The population of near-Earth asteroids (NEAs) shows a large variety of objects in terms of physical and dynamical properties. They are subject to planetary encounters and to strong solar wind and radiation effects. Their study is also motivated by practical reasons regarding space exploration and long-term probability of impact with the Earth.
}
{
  We aim to spectrally characterize a significant sample of NEAs with sizes in the range of $\sim$0.25 - 5.5 km (categorized as large), and search for  connections between their spectral types and the orbital parameters.
}
{
  Optical spectra of NEAs were obtained using the Isaac Newton Telescope (INT) equipped with the IDS spectrograph. These observations are analyzed using taxonomic classification and by comparison with laboratory spectra of meteorites. 
}
{
  A total number of 76 NEAs were observed. We spectrally classified  44 of them as Q/S-complex, 16 as B/C-complex, eight as V-types, and another eight belong to the remaining taxonomic classes. Our sample contains 27 asteroids categorized as potentially hazardous and 31 possible targets for space missions including  (459872) 2014 EK24, (436724) 2011 UW158, and (67367) 2000 LY27. The spectral data corresponding to (276049) 2002 CE26  and (385186) 1994 AW1 shows the 0.7 $\mu$m feature which indicates the presence of hydrated minerals on their surface.  We report that Q-types have the lowest perihelia (a median value and absolute deviation of  $0.797\pm0.244$ AU) and are systematically larger than the S-type asteroids observed in our sample. We explain these observational evidences by thermal fatigue fragmentation as the main process for the rejuvenation of NEA surfaces.
}
{
 In general terms, the taxonomic distribution of our sample is similar to the previous studies and matches the broad groups of the inner main belt asteroids. Nevertheless, we found a wide diversity of spectra compared to the standard taxonomic types.
}

\keywords{minor planets, asteroids  techniques: spectroscopic  methods: observations}

\titlerunning{Near-Earth asteroids spectroscopic survey with Isaac Newton Telescope}
\authorrunning{M. Popescu et al.}
\maketitle
%

\section{Introduction}

The study of near-Earth objects (NEOs) is driven by both scientific reasons and practical motivation. They owe their origin in the main asteroid belt or in the comets population. Some of them are the remnants of the planetesimals that once formed the planets. In this context, the NEAs represent great opportunities to study the origin of our Solar system, tracing the pristine conditions of planetary formation untamed by the influence of major planets and their atmospheres. Because of their proximity to our planet, they provide valuable information about the delivery of water and organic-rich material to the early Earth, and the subsequent emergence of life \citep{2013ApJ...767...54I, 2016E&PSL.441...91M}.

The dynamical and physical characterization of NEOs is important for assessing the orbital evolution of the small bodies in the context of gravitational perturbations by major planets and other subtle phenomena like the Yarkovsky-O'Keefe-Radzievskii-Paddack (YORP), Yarkovsky and space weathering effects \citep{2015aste.book..243B,2015aste.book..509V}. 

The study of NEOs is critical due to their potential threat to Earth (at centuries to geologic timescales). This is evidenced by past impact events like recent super-bolides, known craters and mass extinctions probably caused by major impacts \citep{kel53,1980Sci...208.1095A}. In this sense, improving the physical knowledge about NEOs is essential for planning space-missions aimed to mitigate preventive actions against potential impacts \citep[e.g.][]{2012ExA....33..645B, 2017LPI....48.2652A, 2018P&SS..157..104C, 2018cosp...42E2280M}. 

The potentially hazardous asteroids (PHAs) are those NEAs with an absolute magnitude $H\leq 22$ and the minimum orbit intersection distance (MOID) to Earth orbit smaller than 0.05 AU.  Currently, about 10$\%$ of the known NEAs are classified into this category. \citet{2012ApJ...752..110M} constrained the number of PHAs larger than 100 m to $\sim$ 4700 $\pm$ 1450 objects, by extrapolating and de-biasing the sample of objects detected by WISE. These celestial bodies give a long term risk for colliding with our planet and causing an extensive damage. The "mitigation strategy" is dependent on our ability to determine their physical properties \citep{2016AJ....151...11P}. 

The asteroids that require a low speed budget (denoted as $\Delta V$, and measured in km/s) to be reached  by spacecraft are the most appealing from the point of view of space exploration. The typical requirements are a $\Delta V \leq 7$ km/s for a rendez-vous mission and a $\Delta V \leq 6$ km/s for a sample-return mission \citep{2004MPS...39..351B}. For example, the required $\Delta V$ to reach the Moon and Mars are 6.0 and 6.3 km/s respectively.  At the time of writing, NASA OSIRIS-REx \citep{Lauretta:2017aa} and JAXA Haybausa2 missions \citep{2015aste.book..397Y} reached their targets and will return samples back to Earth from two primitive NEAs (Bennu and Ryugu). These are supposed to have carbon-bearing and organic rich compositions. In addition, primitive NEAs are considered ideal targets for in-situ resource utilization and could become a source of materials for space activities in the near future \citep{2013CeMDA.116..367G, 2014AcAau..96..227E, 2016A&G....57d4.32C}. The Near-Earth Object Human Spaceflight Accessible Targets Study (NHATS) is one of the main initiatives to identify targets of interest. The NHATS lists 1\,382 targets \footnote{\url{http://neo.jpl.nasa.gov/nhats/}}, but less than 5$\%$ have a compositional characterization. 

The proximity of these celestial objects allows Earth observers to study asteroids with sizes ranging from meters to kilometers and to use a variety of observational techniques (e.g. radar, spectroscopy, polarimetry). This population covers objects with about three orders of magnitude smaller than those observable in the main belt. Preliminary results, based only on small sets, have already evidenced that small asteroids behave differently to larger bodies in terms of rotational properties \citep{2013Icar..225..141S}, regolith generation \citep{2014Natur.508..233D} and compositional distributions \citep{2018P&SS..157...82P, 2018MNRAS.476.4481B, 2018MNRAS.477.2786P}. 

About 15 $\%$ of NEAs have  some physical properties determined in the literature (i.e. lightcurves, spectra, colors). This ratio is strongly correlated with their size. The smaller asteroids are difficult to observe, and their observation is opportunistic rather than planned with long time in advance. They become bright enough to be physically characterized from Earth only for very limited time spans, coinciding with their close approaches with our planet, typically during discovery apparition. 

Photometry and spectroscopy are the most used techniques for determining the main physical properties of NEAs such as their size, shape, taxonomy and morphology. Among the largest spectroscopic asteroid surveys were the SMASS programs\footnote{\url{http://smass.mit.edu/smass.html}} \citep{1995Icar..115....1X, 2002Icar..158..106B, 2002Icar..159..468B}. These programs and their continuation, the MIT-UH-IRTF NEO Reconnaissance \footnote{\url{http://smass.mit.edu/minus.html}}, are among the most important spectral data contributors \citep{2019Icar..324...41B}. In this article we will refer to these surveys as SMASS-MIT. They provide preliminary and published spectral results over the visible and near-infrared (NIR) wavelengths for more than one thousand NEAs. \citet{2014Icar..228..217T} used part of this data and their own survey in order to correlate the spectral information with albedos and diameters derived from thermal infrared observations. Based on these findings they were able to distinguish various mineralogies.

Two other dedicated programs which covered a significant sample of NEOs  include the Spectroscopic Investigation of Near Earth Objects (SINEO) which used the ESO New Technology Telescope and the Italian Telescopio Nazionale Galileo (TNG) in order to obtain more than 150 spectra \citep{2005MNRAS.359.1575L, 2008MSAIS..12...20L}; and  the Near-Earth and Mars-Crosser Asteroids Spectroscopic Survey (NEMCASS) which observed 74 asteroids with the Nordic Optical Telescope and the TNG \citep{2010AA...517A..23D}.  Other different groups have reported spectral data for NEAs. Some of these are \citet{2001PhDT.......121W, 2002A&A...391..757A, 2011A&A...535A..15P, 2013Icar..225..131S, 2014PASJ...66...51K, 2014A&A...569A..59I, 2015A&A...584A..97T, 2018arXiv181003706H}.

In the framework of the NEOShield-2 project an extensive observational campaign was performed involving complementary techniques for providing physical and compositional characterization of a large number of NEOs in the hundred-meter size range. They obtained new spectra of NEAs for 147 individual objects \citep{2018P&SS..157...82P}. This number includes 29 asteroids with diameters smaller than 100 meters,  71 objects in the range of 100 m to 300 m, and 47 with sizes up to 650 m. They were the first to provide a comprehensive characterization of a significant sample of NEAs with diameters in the range of ten to hundred meters. 

The Mission Accessible Near-Earth Objects Survey (MANOS) aims at characterizing sub-km, low delta-V (typically $\leq$7 km/s) NEOs by collecting astrometry, lightcurve photometry \citep{2016AJ....152..163T}, and reflectance spectra \citep[e.g.][]{2018DPS....5050801D}. They obtained preliminary data for more than 300 asteroids with a mean size around 80 meters ($H$ = 25 mag), and with some targets as small as few meters.

Broadband photometry is a promising field for rapid characterization of NEAs. It allows preliminary taxonomic classification based on color-color diagrams. The accuracy of this classification is strongly dependent on the photometric errors and calibrations. It can be less precise even compared with spectra that have low signal to noise ratio (SNR). In order to improve the results, certain filters which sample the critical spectral features (like the 1 and 2 $\mu$m bands) should be taken into account for observations. On the other hand, the benefits of classifying an asteroid based on spectrophotometric data are: it requires less observing time, faint targets (i.e., small objects or those at large heliocentric distances) can be characterized, and big datasets are available from sky surveys. Some of the relevant results in this field were reported by \citet{2016AJ....151...98M, 2017AJ....154..162E, 2018ASSP...51...27K, 2018A&A...615A.127I}. By using the Sloan Digital Sky Survey data, \citet{2016Icar..268..340C} provided taxonomic classification for 982 NEAs and Mars-Crossers asteroids. 

In the framework of the EURONEAR~\footnote{\url{www.euronear.org}} collaboration we performed a spectroscopic survey of asteroids using the INT instrument. We called it NAVI (Near-Earth  Asteroids Visible spectroscopic survey with the INT). The aim was to obtain visible spectral data for a statistically significant number of NEAs with previously unknown spectral properties. The observed sample is robust and reliable because a single instrument was used and the same data reduction procedures were applied. This result allows us to derive accurate statistics and to search for connections between spectral properties and orbital parameters. We targeted the asteroids in the 0.25 - 5.5 km size range. This complements the recent NEOShield-2 and MANOS surveys which, in principle, aim to characterize smaller objects. 

The paper is organized as follows: in Section 2 we describe the planning of the observations, the telescope used and the data reduction procedures. The Section 3 shows the methods used to analyze the data. The results are presented in Section 4. The last section is devoted to Discussions and Conclusions. 

\section{Observations and data reduction}

The observational program was performed over four semesters, between 2014A and 2015B. It was divided in 22 observing sessions spread at almost regular time intervals (one night/month).

\subsection{The observing facility}

We used the 2.5 m INT telescope which is owned by the Isaac Newton Group (ING). It is located at an altitude of 2\,336~m at  Observatorio del Roque de los Muchachos in La Palma (Canary Islands, Spain). The Intermediate Dispersion Spectrograph (IDS) is mounted at F/15 Cassegrain focus and it is fed by a long-slit with adjustable width. A set of 16 gratings allows the selection of a resolution ranging from 500 to 9000 (in 1x1 pixel binning of the cameras). The median site seeing is $0.8^{\prime\prime}$, thus we used $1.5^{\prime\prime}$ slit width most of the time (a few exceptions were made with $1.2^{\prime\prime}$ slit width). The INT is capable of tracking at differential rates, and this mode was used to follow all the targets (we carefully monitored and manually adjusted the asteroids in the slit to account for tracking errors).

Two charge-coupled device (CCD) detectors can be used with the IDS. The new \texttt{Red+2} camera (available since 2011) has a size of $2148\times4200$  pixels. The size of each pixel is 15.0 $\mu$m which is equivalent to a scale of $0.44^{\prime\prime}$/pixel. For this camera the fringing is negligible up to 1 $\mu$m. The Red+2 camera was used for the majority of our observations. The \texttt{EEV10} was used to observe only four objects (39796, 52750, 276049, and 436775), being the only option during the respective nights. It has a chip of $2048\times4200$ square pixels of 13.5 $\mu$m width of each. This is equivalent to $0.40^{\prime\prime}$/pixel. The high frequency fringing of this camera decreases the SNR in th red part. To circumvent this effect we binned the data and cut the spectral data above 0.85 $\mu m$  for the corresponding targets.

To observe most of the targets, we used the low resolution grating $R150V$ (maximum efficiency 65$\%$ around $0.5~\mu$m) with central wavelength 0.65 $\mu$m. The $R300V$ grating was used only on the first observing night (2014 February, 11). This configuration covers the spectral interval 0.4-1.0 $\mu$m with a resolution of $R\sim1000$.

The observations were conducted by orienting the slit along the parallactic angle. This procedure minimizes the effects of atmospheric differential refraction. The strategy was to observe most of the targets as close as possible to the zenith . At least one G2V solar analog was observed in the apparent vicinity (similar airmass) of each observed asteroid. We acquired 3-5 image exposures for each target. This avoids the observational artifacts such as the asteroid passing in front of a star, sky variability, the eventual errors of centering of the object in the slit, and cosmic rays. The observational circumstances are shown in Table ~\ref{Circumstances}.

With a history of more than 50 years, the INT continues to remain extremely productive being one of the first 2 m class telescopes. Part of its success, is owned by the ING studentship program which provides INT hands-on training and support every year to 4-6 students. They were invited to take part in our survey by performing the observations. The students were assisted remotely from Astronomical Institute in Bucharest using Virtual Network Connection (VNC). Since February 2015, all observations presented in this paper have been taken following this approach, while all the others were taken in visitor mode.

\begin{figure*}
\begin{center}
\includegraphics[width=8cm]{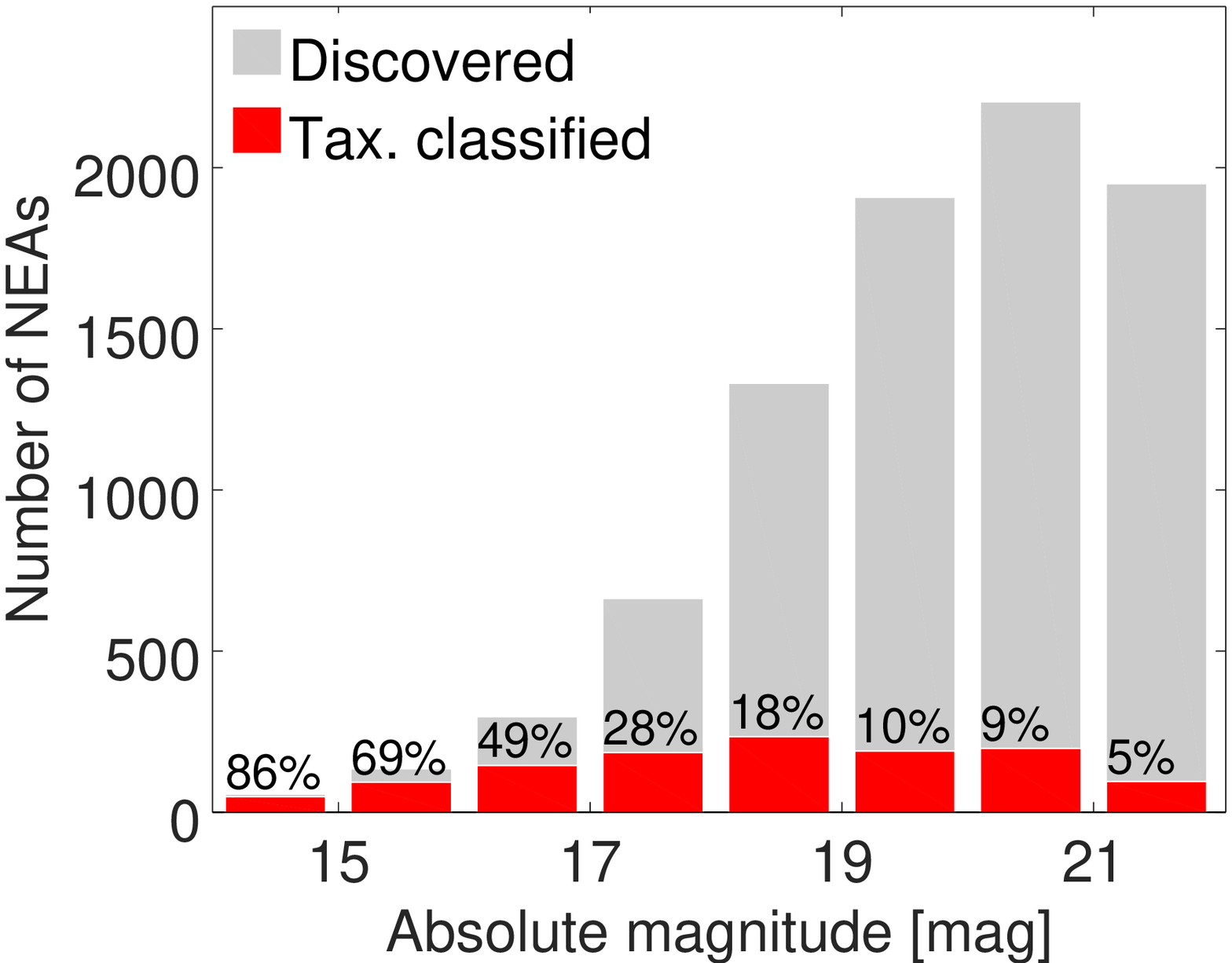}
\includegraphics[width=8cm]{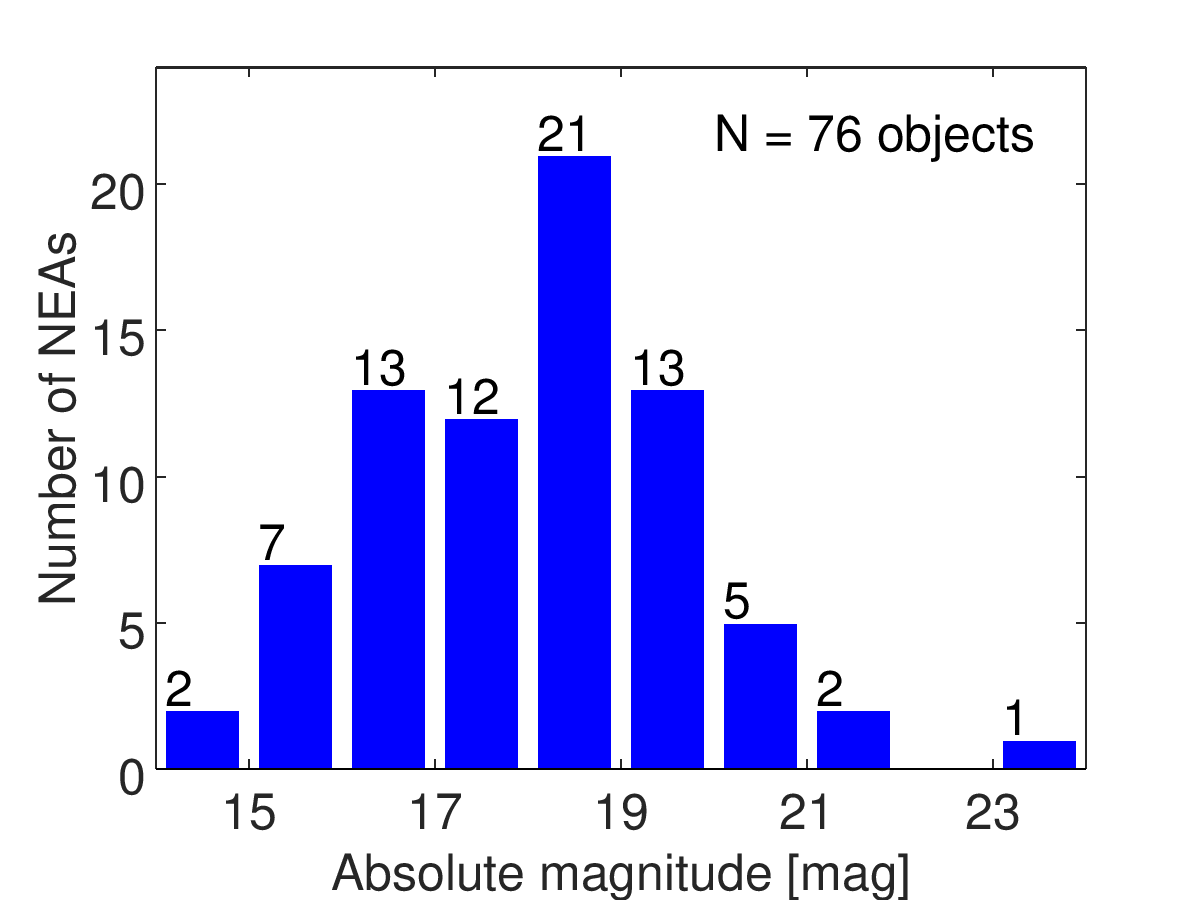}
\end{center}
\caption{(Left) The number of NEAs as  a function of their absolute magnitudes. The discovered objects are shown with grey, while those with a taxonomic classification are shown with red (including this work). The number above each bar represents the percentage of asteroids with taxonomic classification with respect to the discovered ones. (Right) Absolute magnitude distribution of NEAs observed with INT/IDS instrument.}
\label{neasize}
\end{figure*}

\subsection{Sample selection}

The INT/IDS spectrograph allows observations of targets with an apparent magnitude of $m_V \lesssim$ 18.5. Thus, the most accessible NEAs for our program were in the size range of 0.25 - 5.5 km (at the moment of observations the diameter was roughly estimated assuming a typical visual geometric albedo $p_V=0.15$). The histogram of absolute magnitudes for the observed asteroids is shown in Fig.~\ref{neasize} -- right. This is compared with other data available in the literature (Fig.~\ref{neasize} - right).

In order to select the candidates we considered the latest information available on the Minor Planet Center website and we verified if there was any spectral information available on SMASS-MIT and European Asteroid Research Node -- EARN (this website has not been available since $\sim$ 2018, but its information is now included in SSA-NEO web portal \footnote{\url{http://neo.ssa.esa.int/service-description}} ) websites. The purpose was to complement, but not to duplicate the existing data. The next step was to make the planning based on the observational constraints. This requires that each asteroid reaches an altitude above horizon of at least $30^\circ$ (equivalent to an airmass smaller than 2), that its apparent magnitude is brighter than $m_V=18.5$, that the differential rate being slower than 15 $^{\prime\prime}/min$, and that its apparent distance from the Moon being greater than $\sim25^\circ$. We avoided observing objects in crowded fields (i.e. close to the galactic plane). There were $\sim$20-60 targets available on each observing night which met the above criteria. We prioritized the space-mission candidates and the asteroids labeled as PHAs.

In order to confirm our setup we observed several asteroids for which the visible spectra were already known. These are (7889) 1994 LX, (152679) 1998 KU2, and (337866) 2001 WL15. Part of our targets were also observed over the 0.8-2.5 $\mu$m spectral range and reported by MIT-UH-IRTF Joint Campaign for NEO Spectral Reconnaissance. To ensure the reliability of the data we twice observed four objects of different spectrally classes, namely (25916) 2001 CP44 which is classified as Sq, (90075) 2002 VU94 -- a Q-type, (276049) 2002 CE26 -- a Cg-type, and (348400) 2005 JF21 which is V-type. The two observations (Fig.~\ref{doublespec}) gave similar results, within the level of errors, and the one with the best SNR was selected for the analysis.

\subsection{Data Reduction}

The data reduction followed the standard procedures. This included bias subtraction, flat field correction, extraction of two dimensional spectra to one dimensional spectra and wavelength mapping. 

Calibration images (biases, flats, arcs) were obtained at the beginning and at the end of the each observing night. The wavelength map was determined using the emission lines from Copper-Argon and Copper-Neon lamps. The arc charts\footnote{\url{http://www.ing.iac.es/astronomy/observing/manuals/ps/tech_notes/tn133.pdf}} cover 0.305 - 0.980 $\mu$m.

The GNU Octave software package \citep{octave} was used to create scripts for IRAF - Image Reduction and Analysis Facility \citep{1986SPIE..627..733T} to perform these tasks automatically. The extraction of the raw spectrum from the images was made with the \emph{ apall} package. Each image was visually inspected to avoid artifacts such as contamination due to background stars, target tracking errors, or spurious reflections (in case of targets located in the apparent vicinity of the Moon). 

\begin{figure*}
\begin{center}
\includegraphics[width=16cm]{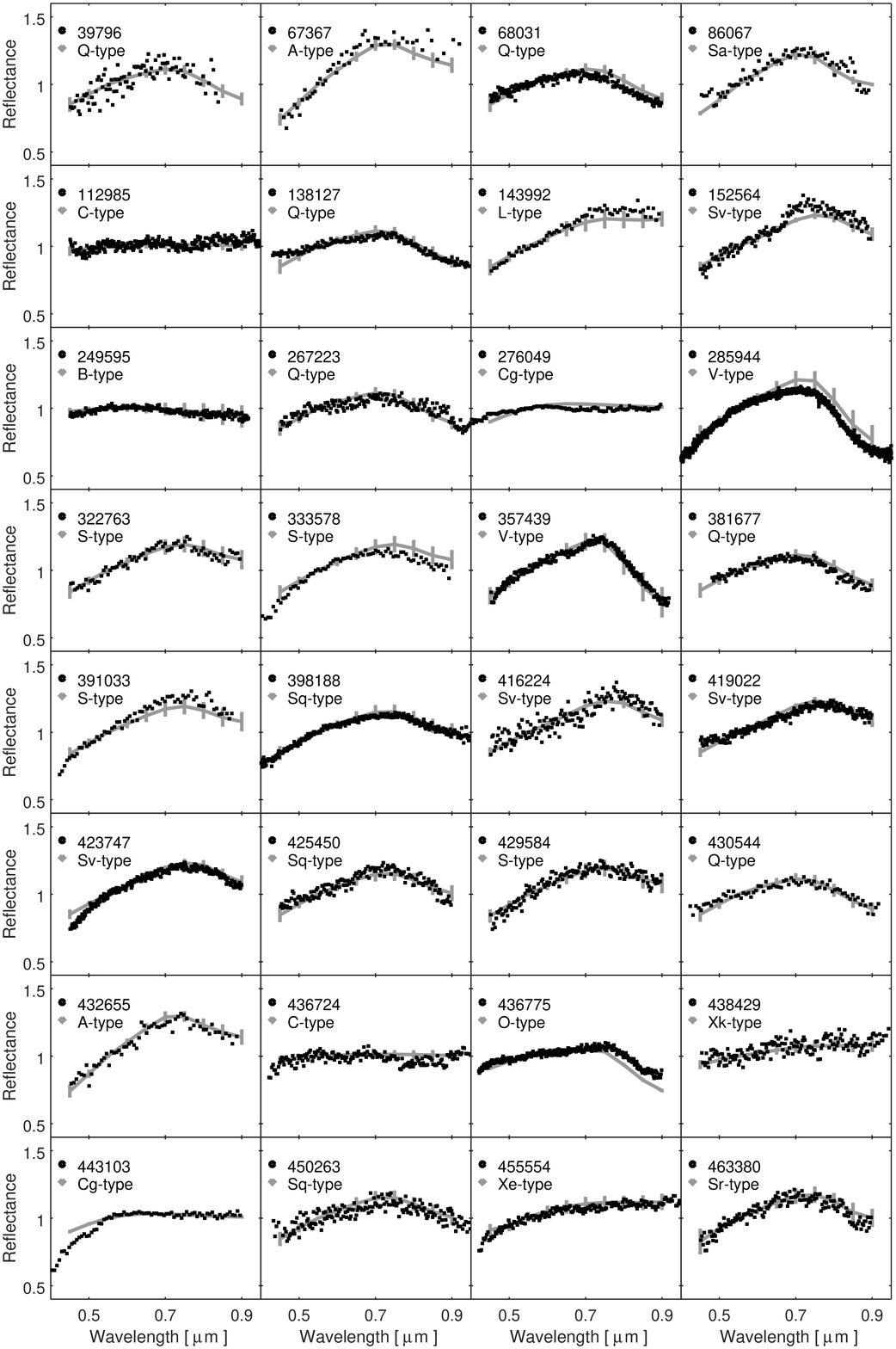}
\end{center}
\end{figure*}
\begin{figure}
\caption{\emph{continuing...}}
\ContinuedFloat
\begin{center}
\includegraphics[width=8cm]{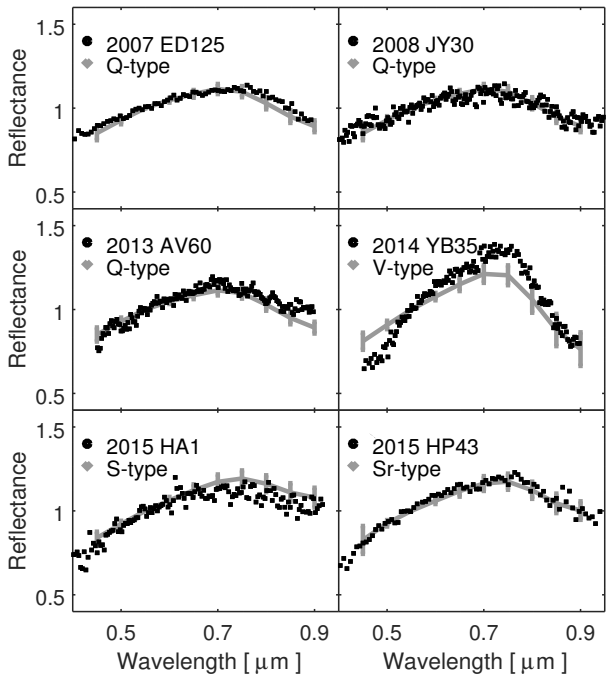}
\end{center}
\caption{The optical spectra of NEAs observed with the INT telescope without NIR counterpart. All reflectances are normalized at 0.55 $\mu$m wavelength. The template spectrum of the assigned taxonomic type is shown in grey. }
\label{VisSpectra}
\end{figure}
%
\begin{figure*}
\begin{center}
\includegraphics[width=16cm]{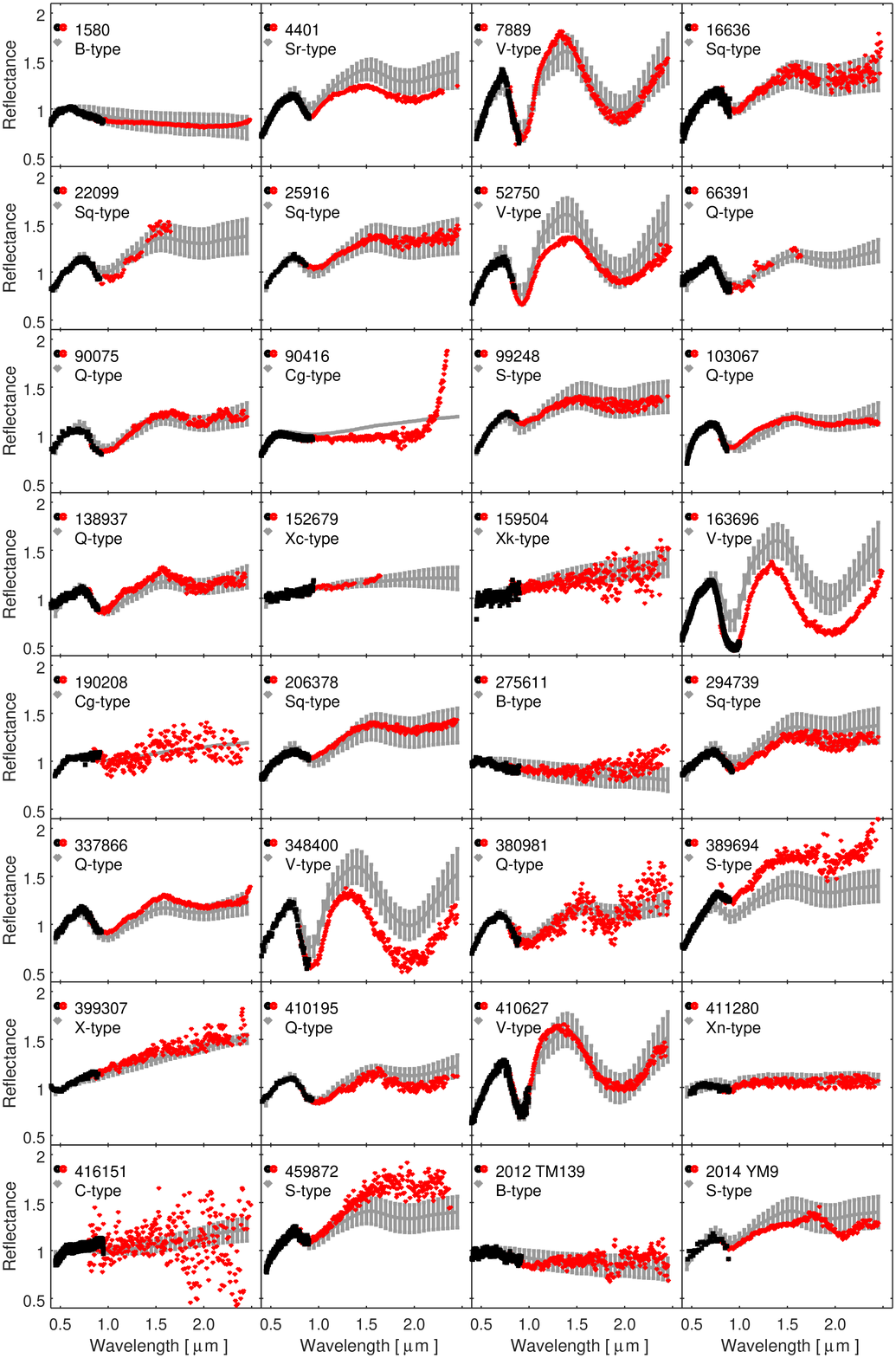}
\end{center}
\end{figure*}
\begin{figure}
\caption{\emph{continuing...}}
\ContinuedFloat
\begin{center}
\includegraphics[width=5.2cm]{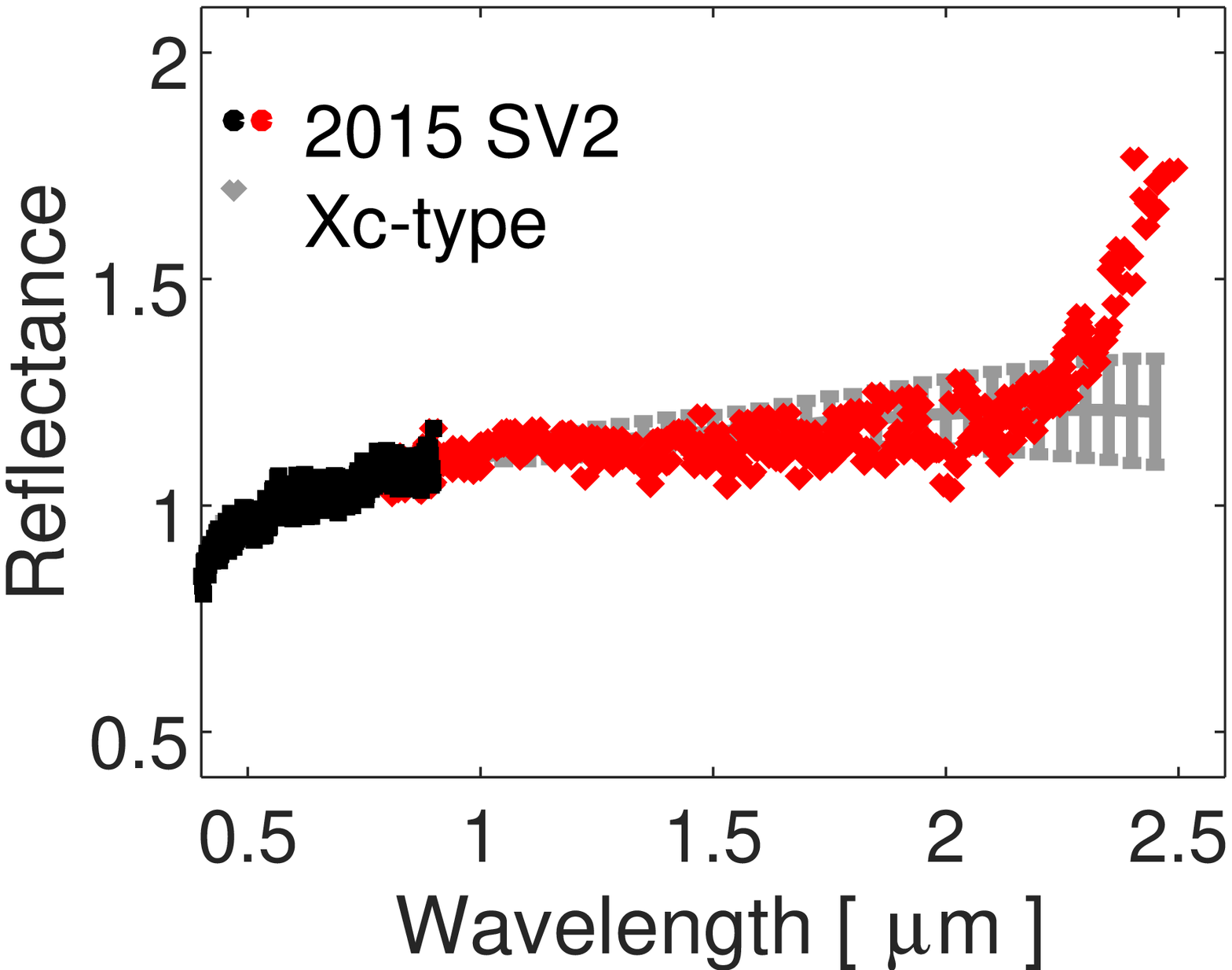}
\end{center}
\caption{The spectra observed with the INT (black points) merged with their NIR counterpart (red points) retrieved from SMASS-MIT website. The obtained spectral curve is normalized at 0.55 $\mu$m wavelength. The template spectrum of the assigned taxonomic type is shown in grey.}
\label{VNIRSpectra}
\end{figure}
\begin{figure}
\begin{center}
\includegraphics[width=8cm]{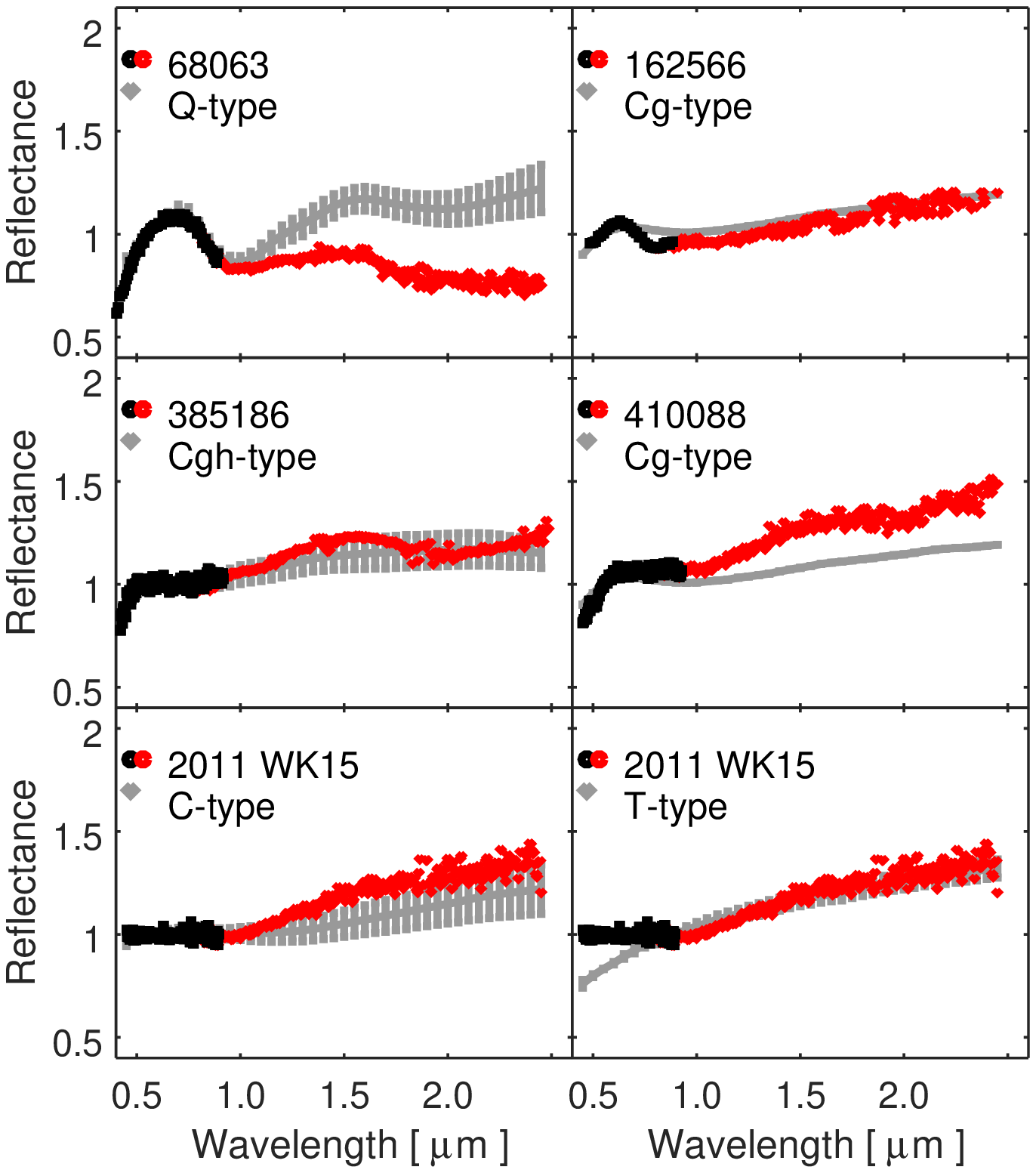}
\end{center}
\caption{Uncommon VNIR spectra. The same format as in Fig.~\ref{VNIRSpectra} is used. The spectrum of 2011 WK15 is compared with C-type assigned based on visible data and with T type assigned based on NIR data.}
\label{unusualspec}
\end{figure}

The spectral reflectance was obtained by dividing the observed asteroid data by a spectrum corresponding to a solar analog. The G2V stars were selected from SIMBAD Astronomical Database - CDS Strasbourg\footnote{\url{http://simbad.u-strasbg.fr/simbad/}}. Several well known solar analogs reported by SMASS surveys were observed for reference and used to verify the solar analogs selection. The possible wavelength shifts between the asteroid and the solar analog (the so called "heartbeats") were corrected by an additional routine.

A binning procedure of four or seven pixels was applied for the spectra with low SNR. For faint objects and for those observed  in poor atmospheric conditions, we shortened the spectral range by cutting further the intervals where the SNR was bellow $\sim$ 3 (due to instrument sensitivity and known telluric absorptions).

All asteroid reflectances were normalized to unity at 0.55 $\mu$m. They are plotted in Fig.~\ref{VisSpectra}, in Fig.~\ref{VNIRSpectra} (here are shown those with NIR counterpart available on SMASS-MIT website), and in Fig.~\ref{unusualspec} (here are shown those with uncommon characteristics).

\section{Methods used to analyze data}\label{MET}

The spectral data in the visible region gives information regarding the surface composition properties of the Solar System airless bodies. The first step in interpreting it is the taxonomic classification. This serves as an alphabet for compositional distribution analysis. Secondly, the size of the asteroids can be estimated based on the correlation between the taxons and the albedo. The next step is the comparison with laboratory data. The mineralogical analysis can be performed for observations covering the visible to NIR wavelengths. By considering the large number of observed NEAs, the statistical determinations are made with respect to the orbital parameters.

Our observed spectra are analyzed in the context of the available information for these NEAs. This enhances the results obtained. The following databases were used: SMASS-MIT website, Near-Earth Asteroid Tisserand Parameters\footnote{\url{https://echo.jpl.nasa.gov/~lance/tisserand/tisserand.html}} and Near-Earth Asteroid Delta-V for Spacecraft Rendezvous\footnote{\url{https://echo.jpl.nasa.gov/~lance/delta_v/delta_v.rendezvous.html}} (these computations were performed by Lance A. M. Benner for Jet Propulsion Laboratory, California Institute of Technology), LCDB -- The Asteroid Lightcurve Database\footnote{\url{http://www.minorplanet.info/lightcurvedatabase.html}} \citep{2009Icar..202..134W}, and the 
NEODyS-2 -- Near Earth Objects Dynamic Site\footnote{\url{https://newton.dm.unipi.it/neodys/}} which pointed to EARN database (no longer accessible) maintained by Dr. Gerhard Hahn for German Aerospace Center (DLR).

We complemented the visible spectra with the NIR ones when available on the SMASS-MIT website. This is the case for 33 objects (as November, 2018). The merging algorithm takes into account the common spectral region ($\approx$0.82 -- 1 $\mu$m, depending on the data). It finds a normalization factor applied to the visible part for which the mean square difference between the two spectra in the common region is minimum. The resulting spectrum (denoted as VNIR) can be better classified taxonomically and it allows the application of the mineralogical models. Nevertheless,  several mismatches may appear in the common wavelength region. These are explained by the fact that the asteroid was observed at different epochs and/or using different setups -- see the discussion from \cite{2014A&A...572A.106P}. 

The spectral slope ($BR_{slope}$) is computed over the 0.45 - 0.7 $\mu$m spectral interval. It provides a way to quantify the reddening effects which can be explained by phase angle changes and space-weathering. This slope may also be used to discriminate between the featureless spectra: the X, T and D-types. The $BR_{slope}$ values, with the units of $\%/(0.1~\mu m)$, are shown in Table~\ref{Circumstances}.

The classification was performed in the framework of the Bus-DeMeo taxonomy. This choice was motivated by the fact that we have both optical and VNIR spectra. This taxonomic system contains 25 classes (24 in its initial version) covering 0.45--2.45 $\mu$m spectral interval and was derived using the principal component analysis over a reference set of 371 asteroids \citep{2009Icar..202..160D}. It follows closely the taxonomy of \citet{1984PhDT.........3T} and it preserved most of the 26 classes of Bus taxonomy \citep{2002Icar..158..146B} with an updated definition for part of them.

To classify an asteroid, we computed the Euclidean distance between the observed spectral curve and the standard spectrum for each of the classes. The best matches were visually reviewed to discriminate the features fitting (which sometimes are not optimally determined by the Euclidean distance between spectra), then the most representative type was assigned. For the VNIR spectra the principal components are computed following the algorithm provided by \citet{2009Icar..202..160D} and both methods were considered for classification.

The diameter of each object can be estimated using the well-known formula \citep[e.g.][]{2007Icar..190..250P}, $D = 1329\cdot(p_V)^{-0.5}\cdot10^{-0.2H}$. The results are shown in Table~\ref{physprop}. The absolute magnitudes $H$ provided by the MPC website were used. The visual albedo $p_V$ was retrieved from the EARN database and it is determined by  WISE observations \citep{2011ApJ...741...68M, 2011ApJ...743..156M, 2014ApJ...792...30M}. When $p_V$ was not available, we considered the average value for the corresponding taxonomic class \citep{2011ApJ...741...90M}. 

The comparison of telescopic observations with laboratory spectra of meteorites can provide relevant compositional information for the objects with prominent spectral features. Consequently, we applied this method to the VNIR spectra of Q/S-complex and V-types (which show bands at 1 and 2 $\mu$m). The analysis was performed using M4AST\footnote{\url{http://m4ast.imcce.fr/}} interface  \citep{2012A&A...544A.130P} which operates with more than 2\,500 meteorites spectra provided in Relab database\footnote{\url{http://www.planetary.brown.edu/relab/}} \citep{2004LPI....35.1720P, 2016LPI....47.2058M}. The results provide an indication to a certain types of compositions, but the solution is not unique and certain parameters such as albedo values and extended spectral range can improve it. 

The band centers and the band area ratio of the spectra corresponding to S-complex and V-type asteroids can be correlated with the mafic mineral abundances. An in-depth review of this technique is provided by \citet{2015aste.book...43R}. They provide a flow-chart for mineralogical characterization of A-, S-, and V-type asteroids based upon the work of several authors including \citet{ 1986JGR....9111641C, 2002aste.book..183G, 2007LPI....38.2117B, 2010Icar..208..789D, 2010M&PS...45..123D, 2011P&SS...59..772R}. We can apply part of this algorithm to the S-complex and V-types NEAs observed by us and that have the VNIR spectra after including the SMASS-MIT data.

We computed the wavelengths of the reflectance minima around 1 $\mu$m and 2 $\mu$m ($BI_{min}$ and $BII_{min}$), the band centers (around 1 $\mu$m - denoted as $BIC$, and 2 $\mu$m - denoted $BIIC$), and the band area ratio ($BAR$) which is the ratio of the areas of the second absorption band relative to the first absorption band.  If there is no overall continuous slope in the spectrum, the band center and the band minimum are coincident. Otherwise, a continuum slope in the spectral region of the absorption feature will displace the band center  by an amount related to the slope of the continuum and the shape of the absorption feature, thus it is removed by division \citep{1986JGR....9111641C}. The results are plotted in $BAR$ vs $BIC$ plot which correlates these parameters with different meteorites composition as determined by \citet{1993Icar..106..573G,2001M&PS...36..761B, 2010M&PS...45.1668C,2013Icar..222..273D}. We tie the analysis only to this graph due to the fact that the NIR spectra of SMASS-MIT website are marked as preliminary and some information is missing (i.e. the observing dates of NIR spectra are required to perform temperature and phase angle corrections).

The SMASS MIT-UH-IRTF website (which is the source of the NIR spectra shown in this paper) warns that although they are free to use for any purpose, these data are unpublished and it is possible that reprocessing or re-calibration could be considered in the future. As a consequence we keep this analysis as an indication of the possible surface compositions and we mention that the errors could actually be larger than the ones shown.

\section{Results}

We observed 92 asteroids in the framework of INT/IDS NEAs spectral survey.  In this article we present the 80 spectra obtained for 76  of NEAs (their full designations are shown in Appendix B) having the $T_J>3$. The $T_J$ is the Tisserand parameter with respect to Jupiter. The rest of 16 objects with $T_J\leq 3$,  were discussed by \citet{2017EPSC...11..537P} and \citet{2018A&A...618A.170L}. The $T_J\approx 3$ is a rough limit between the cometary like orbits ($T_J\leq 3$) and the majority of asteroid orbits ($T_J\geq 3$).

The NEAs physical properties determined in this work, and the other information available for them in the asteroids databases are summarized Table~\ref{physprop}. All observed asteroids were taxonomically classified according to the schema described in Section 3. For a simplified overview we convened the taxonomic classes into four major groups. We found 16 B/C-complex NEAs (in this group we included the B, C-complex, and low albedo X-complex), 44 Q/S-complex objects (we clustered the S-complex, the end-members Q-types, and the peculiar A-types), and 8 basalt-like (V-types). The six objects classified as X-complex asteroids (which, depending of their albedo, can represent various compositions), and the two rare L- and O- types were denoted as \emph{miscellaneous}. A similar approach was also considered by \citet{2016AJ....151...11P} for discussing the nature of PHAs. Overall, we report for the first time the taxonomic type for 59 NEAs. The classification assigned by us for the remaining objects is in agreement with the classification published in the literature and available in the EARN database (Table ~\ref{physprop}).

The largest NEAs of this spectral survey are (1580) Betulia and (25916) 2001 CP44. Their estimated diameter is about $\sim$5\,500 m. Although our survey targeted the asteroids with sizes larger than 250 m, five observed objects have estimated diameter bellow this limit. Among them, the smallest one is (459872) 2014 EK24 with a $\sim$60 m size. 

The combined VNIR spectra of five NEAs show unusual spectral features when comparing with the standard types. These are (68063) 2000 YJ66, (162566) 2000 RJ34, (385186) 1994 AW1, (410088) 2007 EJ, and 2011 WK15. We assigned to them the most likely taxonomic type according to the methods shown in Section 3.  Because we excluded observational artifacts or errors, some particular compositions are conjectured by performing a comparison with spectra from Relab database. Nevertheless, these findings rely on a single observation, thus caution has to be taken and new observations are required to validate and to provide new information about their peculiar nature.

\begin{figure*}
\begin{center}
\includegraphics[width=18cm]{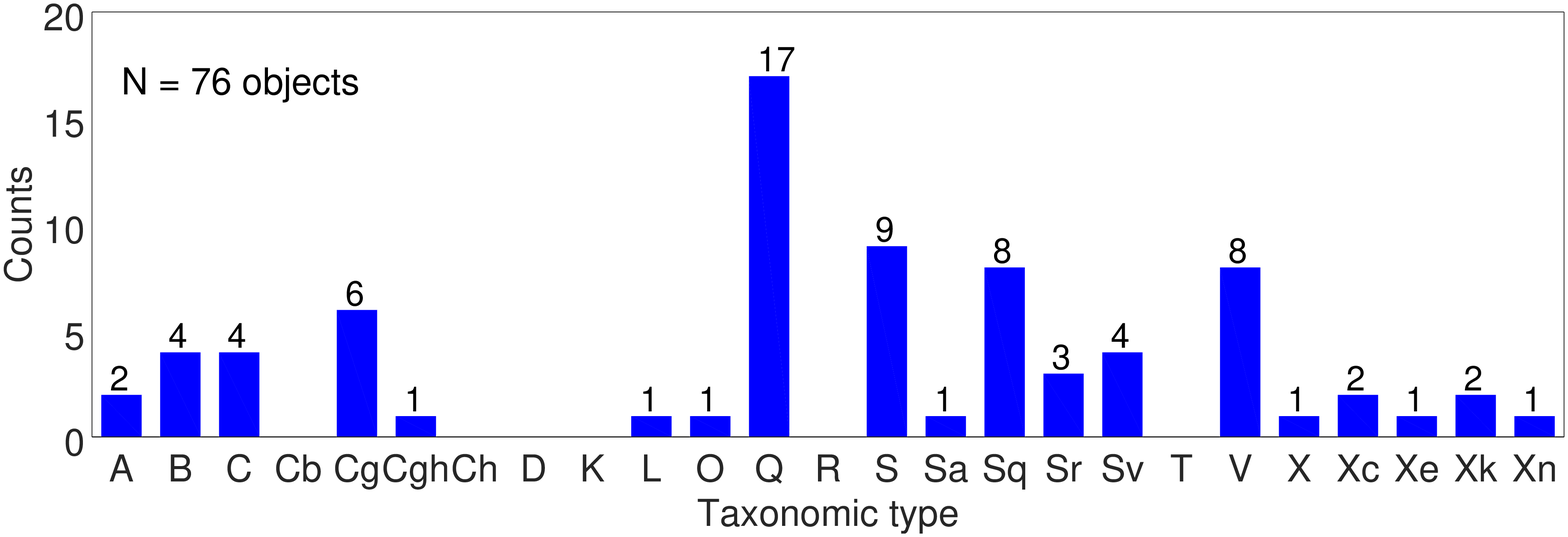}
\end{center}
\caption{The histogram of taxonomic classes for 76 observed NEAs.}
\label{taxclass}
\end{figure*}

\subsection{Taxonomic distributions}

The distribution of the observed NEAs with respect to their classification is shown in Fig.~\ref{taxclass}. The majority of them, $\sim 58\%$ belong to Q/S-complex group. The B/C-complex group covers a fraction of $\sim 21\%$, and the miscellaneous ones are about $\sim 10\%$  from the total number of observed objects. These ratios are in agreement with the ones reported by \citet{2015aste.book..243B}, \citet{2016Icar..268..340C}, and \citet{2019Icar..324...41B}. They found that more than half of all measured NEOs belong to  "S-complex" and the other broad categories, the C- and X-complexes, account for $\sim 15\%$, respectively $\sim 10\%$ of the observations. These broad proportions of spectral types match the taxonomic class distribution of inner main-belt asteroids as reported by \citet{1982Sci...216.1405G} and, more recently, \citet{2014Natur.505..629D}. 

The most represented end-member classes are the V-types (8 objects) and the B-types (4 objects). There are two A-types (olivine rich) asteroids and no D-type was observed.

We note that the relative ratio of  the Q/S-complex and the B/C-complex don't show the same variation with orbital type (i.e.  Amor, Apollo, and Aten which are the name of the most representative member), as the one shown by \citet {2016Icar..268..340C} based on SDSS data who found that more than half of C-complex NEAs are in an Apollo like orbit. We determined that only $25\%$ of the B/C-complex asteroids belong to Apollo dynamic group, while the rest of them ($75\%$) are on an Amor orbit type.

The objects on an Aten like orbit are the most difficult to observe due to their close apparent vicinity to the Sun. Our observations include four asteroids of this kind which we classified spectroscopically as S-complex asteroids (Table~\ref{physprop}).

The distributions reported above are shaped by observational biases. For similar sizes and heliocentric distances, the objects with high albedo (e.g. S-types, V-types) have a brighter apparent magnitude compared to low albedo ones (C-types). Thus, high albedo NEAs (i.e. S-complex, V-types) have a larger probability of being discovered and later to be observed by spectroscopic surveys such as ours. As a consequence,  the low percentage of all classes with low albedo (B/C-complex, D-types) relative to the high albedo ones (Q/S-complex, V-types) is underestimated.

\subsubsection {The B/C-complex group}

This group includes four asteroids classified as B-types, four as C-, six as Cg-, one as Cgh-, and one low albedo Xc- type. They show spectral similarities with carbonaceous chondrite meteorites \citep[e.g.][]{2011Icar..212..180C, 2011Icar..216..309C, 2012Icar..218..196D} and with interplanetary dust particles \citep{2015ApJ...806..204V, 2016A&A...586A..15M}.The albedos are available for five of these objects, and their low values ($p_V \leq 0.1$) are compatible with their taxonomic classification. The diameters of the NEAs belonging to this sample range from 0.4 to 5.5 km.

A classification using the entire VNIR interval was performed for ten of these objects which had NIR data in the SMASS-MIT database. The spectra of (90416) 2003 YK118, (385186) 1994 AW1, (410088) 2007 EJ, and possibly (275611)1999 XX262 show a thermal tail beyond 2 $\mu$m, i.e. they are warm enough to emit detectable thermal flux at these wavelengths \citep{2005Icar..175..175R}.  The VNIR spectra of (162566) 2000 RJ34, and 2011 WK15 show uncommon features, the data in the visible does not seem to the data in the NIR (Fig.~\ref{unusualspec}).

The values of the orbital inclinations suggest a split of B/C-complex in two dynamical subgroups. The first one includes 10 out of 16 observed asteroids with orbital inclinations between 4.1$^\circ$ and 8.3 $^\circ$. These values are compatible with those of low-inclination inner main belt families, Clarissa, Erigone, Polana, and Sulamitis. Their members belong to B/C-complex classes and they represent a likely source for this spectral type of NEAs \citep{2013AJ....146...26C,2016Icar..266...57D,2016A&A...586A.129M,2018A&A...610A..25M}. 
 
The second one is a subgroup of three B/C-complex NEAs, (112985) 2002 RS28, (276049) 2002 CE26, and (1580) Betulia. They show high orbital inclinations (in the range of $47^\circ$ to $52^\circ$). All of them are large objects with diameters in the 2.0 -- 5.5 km range. As a result of their large eccentricity ($e=\sim 0.5$) and orbital inclination, their $T_J$ parameter is close to three. Their dynamical parameters point to a different origin compared to low-inclination ones. The remaining three other NEAs can't be associated to these groups (orbital inclinations in the interval 14--24$^\circ$).

A fraction of 5$\%$ from the total of 76 NEAs observed during this survey are B-types. The percentage is similar to the one found ($\sim$3 $\%$) by \citep{2015aste.book..243B, 2019Icar..324...41B}. The NIR counterpart obtainable from SMASS-MIT for three of these  asteroids, namely 2012 TM139, (275611) 1999 XX262, and (1580) Betulia, shows that the blue slope determined in the visible region is continued over the 1 - 2.45 $\mu$m interval.  We have only the INT/IDS spectrum  for (249595) 1997 GH28. 

The B-types are mostly found in the middle and outer part of the main belt. However, a new result found by \citet{2016Icar..266...57D} showed a significant number of them in the primitive inner main belt families. This taxonomic class resembles the spectra of CV, CO, and CK carbonaceous chondrites \citep{2010JGRE..115.6005C}. Further observations have shown that the NIR spectra of asteroids classified as B-type according to their visible spectra, present a continuous NIR shape variation from negative, blue spectral slopes, to positive red slopes  \citep{2012Icar..218..196D}. This continuum in spectral slopes was also present in the sample of carbonaceous chondrites that best resembled the spectra of B-type asteroids. New findings about these peculiar objects will be reported by NASA OSIRIS-REx mission which has as primary target the B-type near-Earth asteroid Bennu \citep{Lauretta:2017aa}. 

\begin{figure}
\begin{center}
\includegraphics[width=7cm]{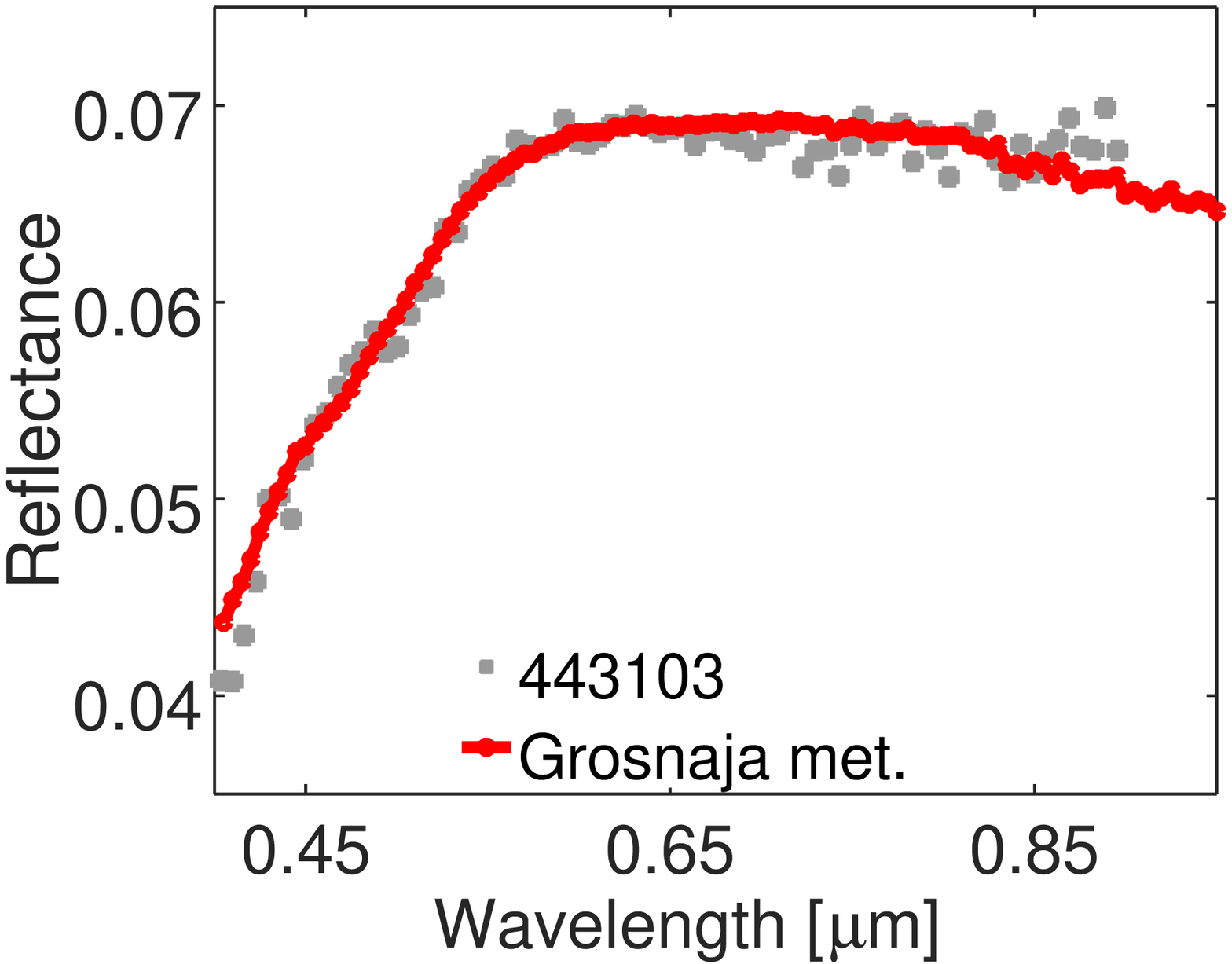}
\end{center}
\caption{The spectrum of (443103) 2013 WT67 in grey  matches the spectrum of CV3 Grosnaja meteorite (Relab sample id RS-CMP-043) in red. The asteroid spectrum was scaled to match the median of reflectance values of the meteorite spectrum.}
\label{443103}
\end{figure}

Six NEAs were classified as Cg and one as Cgh. \citet{2019Icar..324...41B} reported 6 Ch-, 1 Cgh- and 2 Cg-types out of 1040 NEAs. The C-complex asteroids are classified as Cg based on their spectral dropoff before 0.55 $\mu$m \citep{2009Icar..202..160D}, thus a survey covering this interval, as ours, is required to detect them. The Cg type is defined based only on the spectrum of (175) Andromache \citep{2002Icar..158..146B,2009Icar..202..160D}.

The 0.7 $\mu$m absorption feature was detected for one of the spectra (the second spectrum of this object has a lower SNR and this feature is indistinguishable) of (276049) 2002 CE26 and for the spectral data of (385186) 1994 AW1. By using the same procedure as \cite{2016A&A...586A.129M} we measured for the case of (276049) 2002 CE26 a band centered at 0.707 $\pm$ 0.008 $\mu$m with a depth of 1.62$\pm$ 0.23 $\%$, while for (385186) 1994 AW1 there is a band centered at 0.7613 $\pm$ 0.0139 $\mu$m with a depth of 3.68$\pm$ 0.67 $\%$. The associated spectral type is Cgh, where h comes from "hydration". The 0.7 $\mu$m spectral feature is an indication of hydrated minerals on the asteroid surfaces. Their crystalline structure has been altered by their contact with liquid water in the past and over long time-scales. These minerals show their most diagnostic features in the 2.7-2.9 $\mu$m range, the so-called "3 $\mu$m band"\citep{1994Icar..111..456V, 2012Icar..221..744R}. The fraction of Cgh and Ch asteroids in the NEA population is smaller than 1$\%$ \citep[e.g.][]{2015aste.book..243B, 2019Icar..324...41B}. This corresponds with our findings.

An uncommon case is (443103) 2013 WT67. Its spectrum shows a big dropoff in the blue region similar to those of CO/CV carbonaceous chondrite meteorites of petrologic type three. This indicates a primitive nature for this NEA, as these meteorites are the less altered  by thermal and hydration processes. The spectral curve of CV3 Grosnaja meteorite (Relab sample id RS-CMP-043) is one of the best fit to the spectrum of 2013 WT67 (Fig.~\ref{443103}).

No D or T types were identified in this observed sample of 76 asteroids with $T_J \geq 3$. These are rare types in the NEA population. They have a red featureless spectrum which points to an organic and volatile rich composition. The largest known D-type NEAs are (3552), (401857), and (430439) and they are characterized by $T_J \leq 3$. This indicates a possible connection with Jupiter family comets. D-types seem to be more abundant at smaller sizes ($H\geq20$) which is in good agreement with the weakness of the material of this type \citep{2018MNRAS.476.4481B}. Our result provides an upper limit  in the range of 1$\%$ for the presence of these objects with sizes larger than $\sim$250 m.

The carbonaceous like compositions are in general fragile and porous. It is expected that these bodies are gravity-dominated aggregates with negligible tensile strength. By taking into account this fact we can compare the taxonomic classification  with the rotational periods. The spin rates values are available in the LCDB database for all of the objects in the B/C-complex group (Table~\ref{physprop}). All of them are longer than 2.5 hrs, with one exception. This limit is known as cohesionless spin-barrier \citep{2000Icar..148...12P}. Moreover, half of these objects have periods larger than 12 hours, being slow rotators. 

The (436724) 2011 UW158 is a fast rotating asteroid. It has a  rotation period P = $0.610752 \pm 0.000001$ hrs \citep{2016MPBu...43...38C}. This result is incompatible with the fragile carbonaceous like composition. Taking into account its spin state and the featureless spectrum, we can speculate a monolithic body with a possible metallic composition. Its spectral slope (1.9 $\%/\mu$m) computed over 0.45-0.65 $\mu$m is comparable with the lower values of those typical for asteroids recognized as metallic ones \citep{2010Icar..210..655F, 2014Icar..238...37N}.  Additional physical properties (e.g. albedo) are required in order to differentiate between the possible compositions of this object. For computing its diameter, a better guess of albedo is $p_V=0.15$, corresponding to metallic ones \citep{2010Icar..210..655F} which suggest a size in the range of 360 m.

\subsubsection {The Q/S-complex group}

More than half of the observed asteroids are classified as Q-type or S-complex. The SMASS-MIT website contains spectral information over the NIR region for 18 of them. These data confirm and complement our findings. The large majority (40 objects) of NEAs we found in the silicate like group covers the 0.15-2.6 km size range.

For the S-complex asteroids we found 25 objects (9 -S, 1-Sa, 8 – Sq,  3 – Sr and 4- Sv).  Taking into consideration their 1 $\mu m$ band, we also included in this broad group the two A-types which represent olivine rich compositions. The Q-types (17 NEAs) are the most numerous class of this group, representing $21.8\%$ relative to the entire set of observed NEAs. It is followed by the S-type (nine NEAs). Eight NEAs were assigned to Sq- types, which is an intermediate class between Q-types and S-types. The fraction of Q-types is double compared to the one reported by \cite{2019Icar..324...41B} who found that they comprise about $10\%$ of the observed population. The spectral features of this taxonomic class \citep{1985Sci...229..160M} are similar to ordinary chondrite meteorites which are the most abundant.

The orbital types distribution of the Q/S-complex include 20 Amors, 18 Apollos, and 4 Atens. This distribution is roughly the same when compared to the whole known NEA population (as of April 2019 the MPC list the orbits of 8425 Amors -- 42$\%$, 10033 Apollos --50$\%$, and 1522 Atens -- 8 $\%$ ). It outlines the fact that there is no preference for a particular orbit type of the Q/S-complex NEAs. Nevertheless, a  particular subset has large eccentricities and low perihelia. The most noticeable are (267223) 2001 DQ8 ($q$ = 0.183 AU and $e$ = 0.9), (66391) 1999 KW4 ($q$ = 0.2 AU and $e$ = 0.689), and (138127) 2000 EE14 ($q$ = 0.309 AU and $e$ = 0.533). These asteroids are subject to high temperature variations. For example, we mention the extreme case of (267223) 2001 DQ8 which has the surface temperature at perihelion of about 625 K, while at the aphelion ($Q$ = 3.502 AU) the value goes down to 150 K (the temperatures were evaluated following \citet{2015aste.book...43R} and references therein). The large temperature variations may lead to thermal fatigue followed by thermal fragmentation \citep{2014Natur.508..233D}. 

\begin{figure}
\begin{center}
\includegraphics[width=8cm]{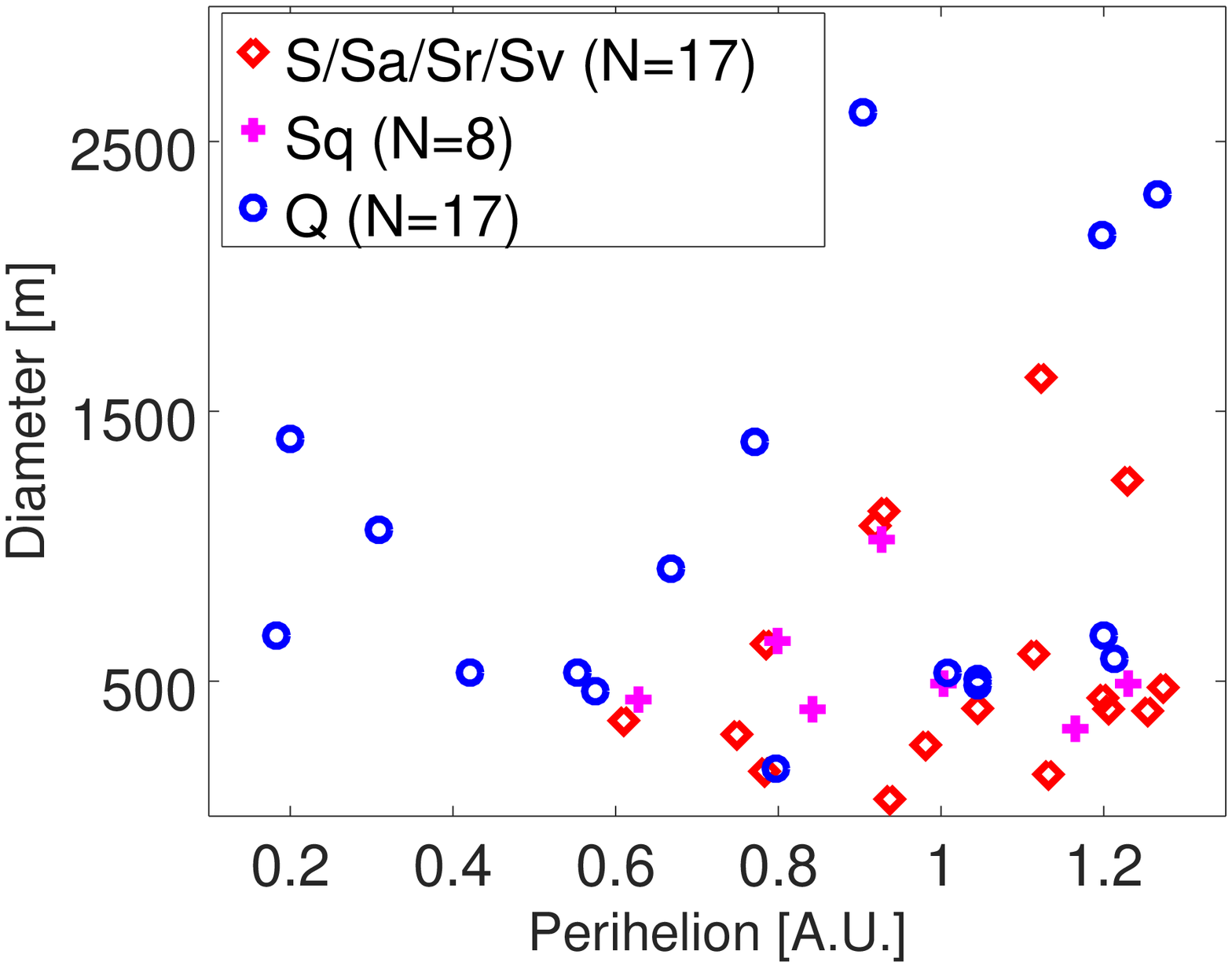}
\end{center}
\caption{The distribution of the Q/S-complex group in the perihelion ($q$) versus diameter space. The Q-type (plotted with blue) asteroids which are considered to have fresh surfaces are shown in comparison with S/Sr/Sv/Sa-types (plotted with red) which have surfaces affected by space-weathering. The Sq type (plotted with magenta) is an intermediate class between S-types and Q-type. For the clarity of the figure, the axis limits do not include Sq type asteroid 25916 (2001 CP44) which has a diameter of 5.6 km and a perihelion of 1.286 A.U.}
\label{diamvsq}
\end{figure}

We found that all low perihelia NEAs ($q~\leq~\sim$0.6 AU) are classified as Q types (Fig.~\ref{diamvsq}). The probability to reach this result by hazard is only 0.0024 (this was computed for our sample considering that Q-types and S-types have equal chance to be observed). This taxonomic class is associated with fresh surfaces, unweathered by processes such as the solar wind sputtering and micrometeorite bombardments \citep{2010Natur.463..331B}. There is spectral continuity between Q and S-complex asteroids can be associated with changes due to space-weathering \citep{1996Sci...273..946B, 2011Sci...333.1113N, 2014Icar..227..112D}. The Q-types can be recognized mostly from the optical part of the spectrum, because the space-weathering effects are more prominent in this region \citep{2014Icar..227..112D}. In the NIR region they are almost identical with the Sq-types and if only this part of the spectrum is used, then accurate spectral data are required to separate these two classes. Within our sample, we complemented the optical part with NIR information from SMASS-MIT for eight Q-types. The initial classification obtained from our observations is confirmed when merging with NIR counterpart.

Various hypothesis have been proposed for the origin of the Q-type resurfaced material. They range from collisions \citep{2004Icar..170..259B} to planetary encounters \citep{2005Icar..173..132N}.  \citet{2010Natur.463..331B} argue that the tidal stress which is strong enough to disturb and expose unweathered surface grain, is the most likely dominant, short-term, S-type asteroid resurfacing process. \cite{2010Icar..209..510N} found that the effect of all terrestrial planetary encounters can therefore explain the tendency towards seeing the Q-types among NEAs. \citet{2014Icar..227..112D} showed the distribution of Q is not random with perihelion, highlighting the effect of Venus and the Earth. However, by using spectro-photometric data from SDSS, \citep{2016Icar..268..340C} showed that Q-type candidates are present among Mars-Crossers, and their encounters with Mars are not enough to explain the resurfacing. \citet{2018Icar..304..162G} presented a model of YORP spin-up that may explain the observations.

The Fig.~\ref{diamvsq} outlines several questions regarding the surface rejuvenating mechanisms proposed by \citet{1985Sci...229..160M, 2004Icar..170..259B, 2005Icar..173..132N, 2010Natur.463..331B}; and \citet{2010Icar..209..510N}. Firstly, why  Q-types show a distributional peak at low perihelion?  At this heliocentric distance, asteroids are expected to have their surfaces strongly affected by space weather (thus, similar to S-complex spectra with red slopes) as the solar wind is stronger and the micrometeorites bombardments are more energetic.

Secondly, why are the Q-types, in general, larger than the S-complex asteroids? The median value of the diameter and its median absolute deviation for the Q-types found in our sample is 670 $\pm$ 140 m, while for the S-complex these values are 440 $\pm$ 220 m (the value is even lower if we exclude the intermediate Sq-types). This finding contradicts the rejuvenating mechanisms due to tidal stress during close encounters with planets, which is more efficient in the case of small bodies due to their low gravity. In this case, the observational bias can be excluded as the average albedo of S-complex and Q-types is similar, $p_V^{average} =~0.22\pm0.07$ \citep{2011ApJ...741...90M, 2011AJ....142...85T}. This is also shown by the albedo values reported in Table~\ref{physprop}. As a consequence, they have the same brightness for the same size and the same probability to be observed.

Furthermore, the low perihelia objects tend to have large diameters. All objects observed by us with a perihelion lower than 0.6 AU are larger than 450 m (as in the case of six NEAs).

One of the processes which can answer the above questions is thermal fracturing \citep{2014Natur.508..233D}. The fragmentation due to thermal fatigue is a rock weathering process which leads to regolith generation. \cite {2014Natur.508..233D} noted that the fresh regolith production by thermal fatigue fragmentation may be an important process for the rejuvenation of NEA surfaces. Our data indicates that this process has to be the most efficient for low perihelion objects. Our results are also in agreement with the prediction that small NEAs could be eroded by thermal fragmentation and radiation pressure sweeping on timescales shorter than their dynamical lifetime \citep{2014Natur.508..233D}.

\citet{2006MNRAS.368L..39M} showed that the Q- and Sq-type planet-crossing asteroids are more abundant at small perihelia (the (Q+Sq)/S ratio has the peak value of 0.9 for perihelion distances around 0.5 AU). By using optical spectra of 214 NEOs obtained by SINEO survey, they found a statistically significant correlation between the spectral slope defined over the visible interval and the perihelion distance. They explained this correlation by the fact that planet-crossing asteroids with small perihelion distances have undergone more frequent planetary encounters than those residing in more distant orbits.

The predominance of Q-types at low perihelia distances is supported also by the results of \citep{2014Icar..227..112D}. They used a sample of 249 S-complex and Q-type NEOs, mostly observed in the NIR spectral region, and found that the Q/S-complex fraction is highest at 0.5 AU (about 0.5 $\%$). \citep{2014Icar..227..112D} also noted that the Q-types are more abundant at smaller perihelia distance. 

Recently, \citet{2019Icar..324...41B} correlated the spectral properties of 195 NEOs falling in the category of Q-, Sq-, and other S-complex types with the perihelion distance. They used the \emph{space weathering parameter} witch takes into account the alteration of spectral features that mimic laboratory measured space weathering alteration effects \citep{2015aste.book..597B}. The low value of this parameter corresponds to Q-types. Their binned results show a clear progression towards decreasing the \emph{space weathering parameter} (thus, suggesting a larger number of Q-types) with decreasing the perihelion distances.  \cite{2019Icar..324...41B} considered the lack of weathered surface inside the perihelion of Venus as a combination of multiple factors, including planetary encounters, YORP spin-up and thermal cycling.

\begin{figure}
\begin{center}
\includegraphics[width=8cm]{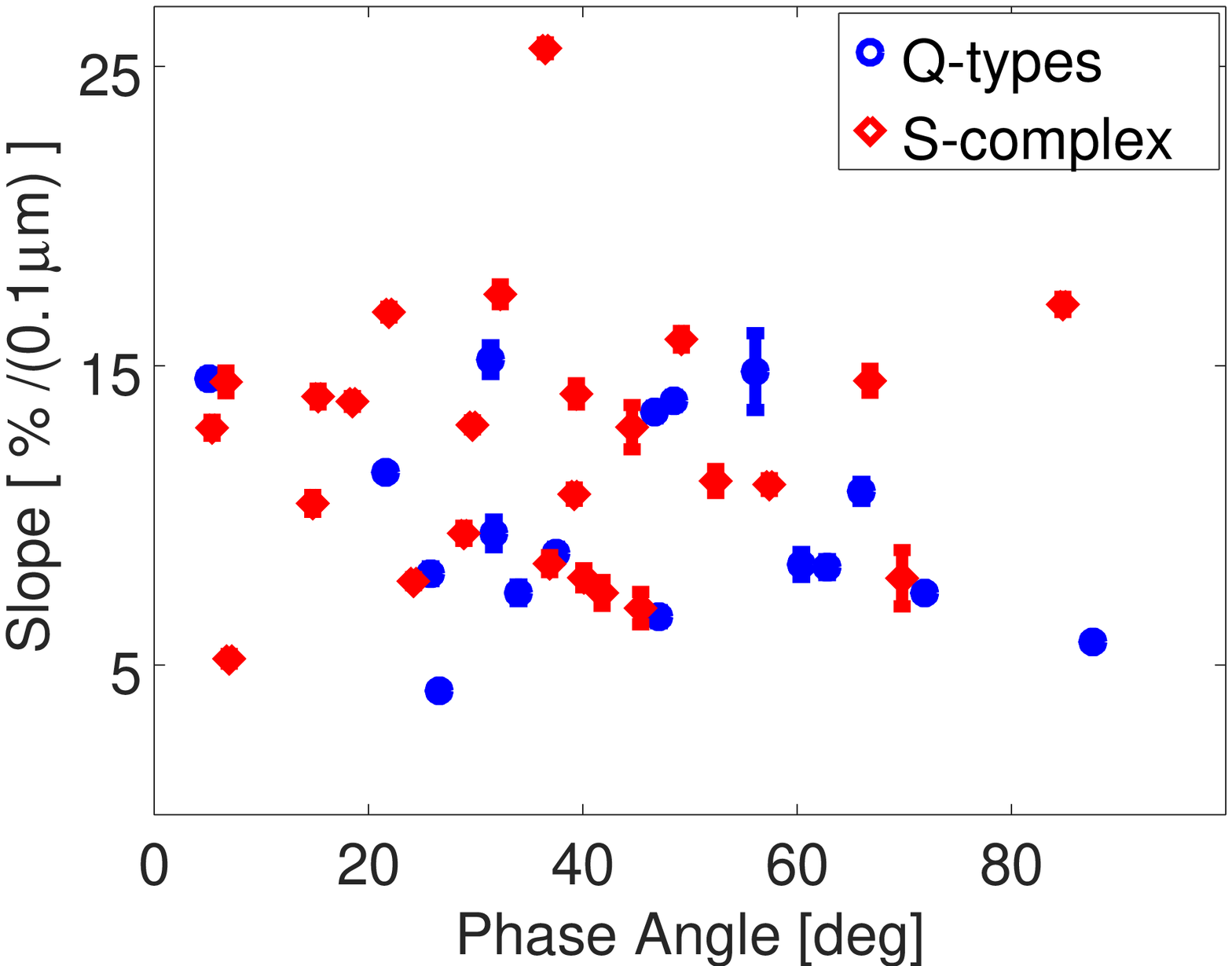}
\end{center}
\caption{ The $BR_{slope}$ versus phase angle for the asteroids belonging to Q/S-complex.}
\label{phasered}
\end{figure}

The phase angle at which the objects are observed influences their spectral slope, a phenomena called phase reddening \citep[][and references therein]{2015aste.book...43R}. Using the SDSS spectro-photometric data, \cite{2015AA...580A..98C} noted that the asteroids  classified  in  most  of  the  major  taxonomic  groups show an increase in spectral slope with increasing phase angle, but there is also a significant number of objects for which these parameters do not follow the correlation. As our observations for the Q and S-complex asteroids were performed over a wide interval of phase angles, we tested if the phase reddening effect can be outlined by plotting the corresponding $BR_{slope}$ of the spectra (Fig.~\ref{phasered}). The result is negative, we are not able to identify a particular trend in the data. The large dispersion of the points suggests that the taxonomic type is a weak parameter to constrain this phenomena and most probably mineralogical variations and surface roughness play a more important role. 

The next step of data analysis is to perform the comparison with Relab meteorites spectra and to compute the band centers and band area ratios for a sub-sample of 14 Q/S-complex NEAs. These were selected for their spectral coverage over 0.45 - 2.45 $\mu$m wavelength interval and their high signal to noise ratio (SNR) data. The results are shown in Table~\ref{Meteorite}. They outline the matching with ordinary chondrites, and in particular with L and LL subtypes with evolved petrologic class (5, 6). This is expected because most of their spectra are classified as Q, or Sq types. The matching with some irradiated chips from Chateau Renard (OC/L6) and Appley Bridge (OC/LL6) meteorites outline the effect of space-weathering.

\begin{table*} 
\caption{Summary of the results obtained by matching the asteroid spectra with those from the Relab database. The asteroid designation and its taxonomic type, the meteorite name, the type (OC -- ordinary chondrites), subtype and petrologic group, the sample ID, the Relab file identifier, and other meteorites types which show spectral matches with the asteroid are provided.}
\centering
\begin{tabular}{l c l l l l l}\hline\hline
Asteroid &Tax. type &Meteorite                  &Type       &SampleID      &RelabFile&Other met. Types  \\ \hline

4401     & Sr       &Homestead                  &OC/L5      &MR-MJG-048    &MGN075   &OC/L5, L4, H3  \\
16636    & Sq       &Chateau Renard, chip irrad.&OC/L6      &OC-TXH-011-A80&C1OC11A80&OC/L6  \\
25916    & Sq       &Hamlet                     &OC/LL4     &OC-TXH-002-B  &C1OC02B  &OC/L4  \\
90075    & Q        &Chateau Renard, chip irrad.&OC/L6      &OC-TXH-011-A60&C1OC11A60&OC/L6, LL5, LL6  \\
99248    & S        &Hamlet                     &OC/LL4     &OC-TXH-002-B  &C1OC02B  &OC/LL4, L4  \\
103067   & Q        &Malakal                    &OC/L5      &TB-TJM-109    &LATB109  &OC/L5, LL4  \\
138937   & Q        &Benares                    &OC/LL4     &MT-HYM-083    &C1MT83   &OC/LL4  \\
206378   & Sq       &Chateau Renard, chip irrad.&OC/L6      &OC-TXH-011-D35&C1OC11D35&OC/L6   \\
294739   & Sq       &Chateau Renard, chip irrad.&OC/L6      &OC-TXH-011-D15&C1OC11D15&OC/L6,LL4  \\
337866   & Q        &Appley Bridge, chip irrad. &OCLL6      &OC-TXH-012-A40&C1OC12A40&OC/LL6,L6  \\
389694   & S        &Y-791058,51                &Stony-Iron &MB-TXH-027    &S2MB27   &Stony-Iron  \\
410195   & Q        &Olivenza                   &OC/LL5     &MT-HYM-085    &C1MT85   &OC/LL5, L6, LL6  \\
459872   & S        &Y-791058,51                &Stony-Iron &MB-TXH-027    &S2MB27   &Stony-Iron, Olivine  \\
2014YM9  & S        &Cynthiana,pellet irrad.    &OC/L4      &OC-TXH-015-D15&C1OC15D15&OC/L4,LL4  \\ \hline
7889     & V        &ALH-78132,61               &Eucrite    &MB-TXH-072-B  &CBMB72   &Eucrites  \\
52750    & V        &EETA79005,99               &Eucrite    &MP-TXH-072-A  &CAMP72   &Eucrites, Howardites  \\
163696   & V        &MIL07001                   &Diogenite  &MT-AWB-168-A  &C1MT168A &Diogenites  \\
348400   & V        &MIL07001                   &Diogenite  &MT-AWB-168-A  &C1MT168A &Diogenites  \\
410627   & V        &Y-790727,144               &Howardite  &MP-TXH-098-A  &CAMP98   &Eucrites  \\
\hline
\end{tabular}
\label{Meteorite}     
\end{table*}

\begin{figure}
\begin{center}
\includegraphics[width=8cm]{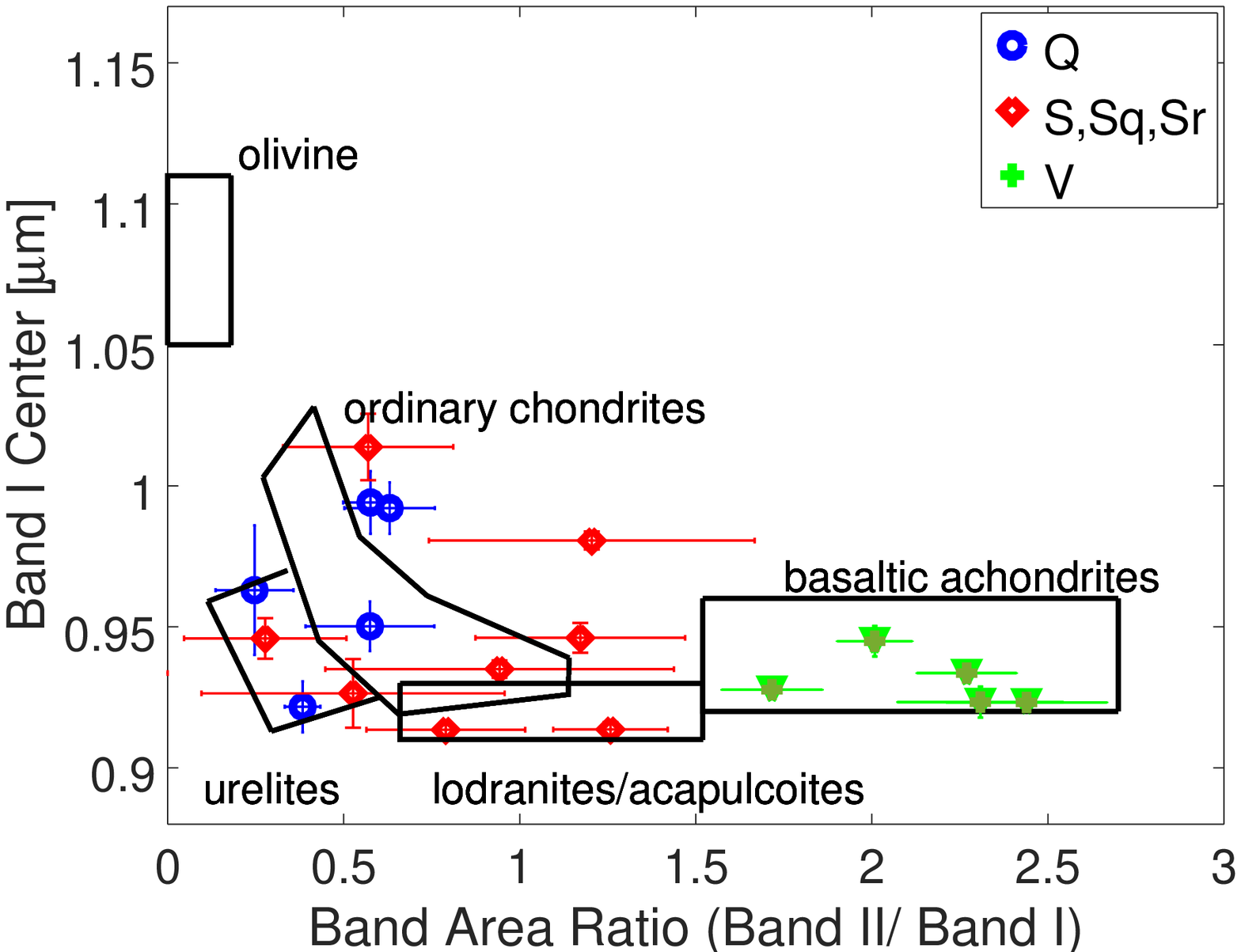}
\end{center}
\caption{The distribution of the observed Q, S-complex and V-types asteroids in the BIC versus BAR space (only data with VNIR spectra are considered). The encircled regions are taken from \cite{2013Icar..222..273D} which follow the results of \cite{1993Icar..106..573G,2001M&PS...36..761B, 2010M&PS...45.1668C}.}
\label{mineralogy}
\end{figure}

Statistically, the first ranked matches with meteorite spectra are in agreement with those found based on BIC and BAR (Fig.~\ref{mineralogy}). The various regions shown in Fig.~\ref{mineralogy} were described by \cite{2013Icar..222..273D} and follow the results of \cite{1993Icar..106..573G} for olivine, ordinary chondrite, and basaltic achondrite regions; \cite{2001M&PS...36..761B} for the primitive achondrite region (lodranites/acapulcoites); and \cite{2010M&PS...45.1668C} for the ureilite region. Most of the Q/S-complex asteroids are placed in the ordinary chondrites region. There are several objects within the ureilites region, but their large errorbars and the fact that we did not apply temperature corrections makes this result uncertain. A particular case is that of (4401) Aditi which shows spectral parameters compatible with the acapulcoites/lodranites primitive achondrites. These meteorites are residues of varying degrees of partial melting with a chondritic composition but with an achondritic texture \citep[e.g.][]{2002cem..book.....N}.

The composition inferred from the spectral data can be compared with the structure expected from the spin rates. The LCDB database contains rotation periods for 37 Q/S-complex NEAs. The fastest rotator is the 65 m S-type asteroid (459872) 2014 EK24. It has a rotation period of 0.0998 $\pm$ 0.0002 hrs (i.e. about 6 min) and a lightcurve amplitude between 0.56 and 1.26 mag \citep{2016MPBu...43..156G, 2016AJ....152..163T, 2018MPBu...45...57T}. All other NEAs from this sample have rotation periods larger than 2.1 hrs. There are also several slow rotators, i.e rotation periods larger than 24 hrs including  (206378) 2003 RB, (429584) 2011 EU29, 2015 HA1, (410195) 2007 RT147, (39796) 1997 TD, (463380) 2013 BY45, and (391033) 2005 TR15. This sample doesn't show any dependencies between the taxonomic type and the rotation period.

\subsubsection {The basaltic like group}

Our observations contain eight NEAs classified as V-types. Their spectra resemble those of howardite--eucrite--diogenite meteorites. It is assumed that they have a similar composition to (4) Vesta which is considered as their parent body. They are also called basaltic asteroids due to their igneous nature. The latest reported spectro-photometric or spectral data \citep[e.g.][]{2016Icar..268..340C, 2016MNRAS.455.2871I, 2014Icar..235...60H, 2010AA...517A..23D, 2006AA...456..775D} shows that this spectral type represents about 10 $\%$ of the NEA population. Within our sample we found a similar fraction (8 out 76 observed asteroids).

The smallest object in this basaltic group is (410627) 2008 RG1 with an equivalent size of about 160 m. Four of the V-types have the equivalent diameter greater than 1 km  and the largest is (7889) 1994 LX which has a size of about 1680 m. In the SMASS - MIT database there are near infrared spectral data for five of them and confirm our classification.

The sample of basaltic NEAs shows a wide spread of orbital inclinations ranging between 10.8$^\circ$ and 53.1$^\circ$. In particular three of the largest basaltic NEAs, (163696) 2003 EB50, (7889) 1994 LX and (285944) 2001 RZ11 (1+ km) show very high orbital inclination (29.5$^\circ$, 36.9$^\circ$, 53.1$^\circ$ respectively). Compared to these values, the Vesta collisional family spans the 5$^\circ$ and 8$^\circ$ orbital inclinations \citep{2015aste.book..297N}. The V-types presented here are on Apollo and Amor type orbits and their eccentricities values range from 0.35 to 0.54. The small minimum orbital intersection distance (MOID) with Earth of four of these objects categorize them as PHAs.

The rotation periods and lightcurve amplitudes are available for seven of the basaltic asteroids. They can be clustered in V-type NEAs with  synodic periods between 2.2 and 3.3 hours, and the slow rotators (52750) 1998 KK17 $P_{syn} = 26.43$ hrs and (163696) 2003 EB50  $P_{syn} = 62.4$ hrs. The majority of them show small lightcurve amplitude (i.e. bellow 0.25) suggesting a close to round shape, with the exception of the slow rotator (163696) 2003 EB50 which has $A_{max}$ = 1.64 mag. This points to a highly elongated object \citep{2016MPBu...43..240W,2017MPBu...44..200O}.

The comparison to meteorites data from Relab shows spectral matches with  HEDs (Table~\ref{Meteorite}).  The NEA (7889) 1994 LX shows a match with the eucrites, (163696) 2003 EB50 and (348400) 2005 JF21 shows a match with diogenites,  while (52750) 1998 KK17 and (410627) 2008 RG1 have similarities with Howardites/eucrites group. These results are supported by their band parameters (Fig.~\ref{mineralogy}).

\subsection{Miscellaneous types}

The miscellaneous group includes an L-type, an O-type and six X-complex objects. The L-types and O-types are end member classes with peculiar spectral properties. The X-complex includes both high and low albedo objects and may represent different compositions. It covers the E (enstatites), M (metallic), and P (primitives) classes from  the \cite{1984PhDT.........3T} definition.

Based only on the INT/IDS visible spectrum we classified (143992) 2004 AF as L-type. This taxonomic class exhibits variations in terms of slope and spectral features mostly in the NIR region. These are interpreted as diagnostic of minerals with a high FeO content \citep{2018Icar..304...31D}. However, based only on the optical region the classification is not uniquely determined as other classes such as S, K, and A can not be totally excluded.

According to its visible spectrum, (436775) 2012 LC1 is the O-type. \citet{2011LPI....42.2483B} noted that these are asteroids with absorption bands similar to pyroxenes but with band minima that are not typically found in terrestrial samples or in the meteorites. The representative member of this class is (3628) Bo\v{z}n\v{e}mcov\'{a} and very few objects with similar spectral properties are known to exist \citep{2011LPI....42.2483B}. Within the NEA population there are several objects marked as O-types, according to \citet{2015aste.book..243B}. We note that some of the objects found initially as O-type NEAs based on their visible spectrum, were reclassified as Q-type or V-types when NIR data were added. One such example is (5143) Heracles which was classified as O-type by \citet{2002Icar..158..146B} and reclassified as Q-type \citep{2009Icar..202..160D, 2014A&A...572A.106P}. 

The NIR data available for four of the X-complex asteroids from SMASS database confirm our classification. The NIR spectrum of asteroid 2015 SV2 recovered from SMASS-MIT database shows a thermal tail. This points to a low albedo object. We note that (399307) 1991 RJ2 shows an absorption feature around 0.49 $\mu$m that resembles the absorption seen on some E-type asteroids  such as (2867) {\v S}teins and (3103) Eger. \citet{2007A&A...474L..29F} noted that these features are possible due to the calcium sulfide mineral oldhamite (present in highly reduced assemblages such as aubrites).

\begin{figure*}
\begin{center}
\includegraphics[width=4.5cm]{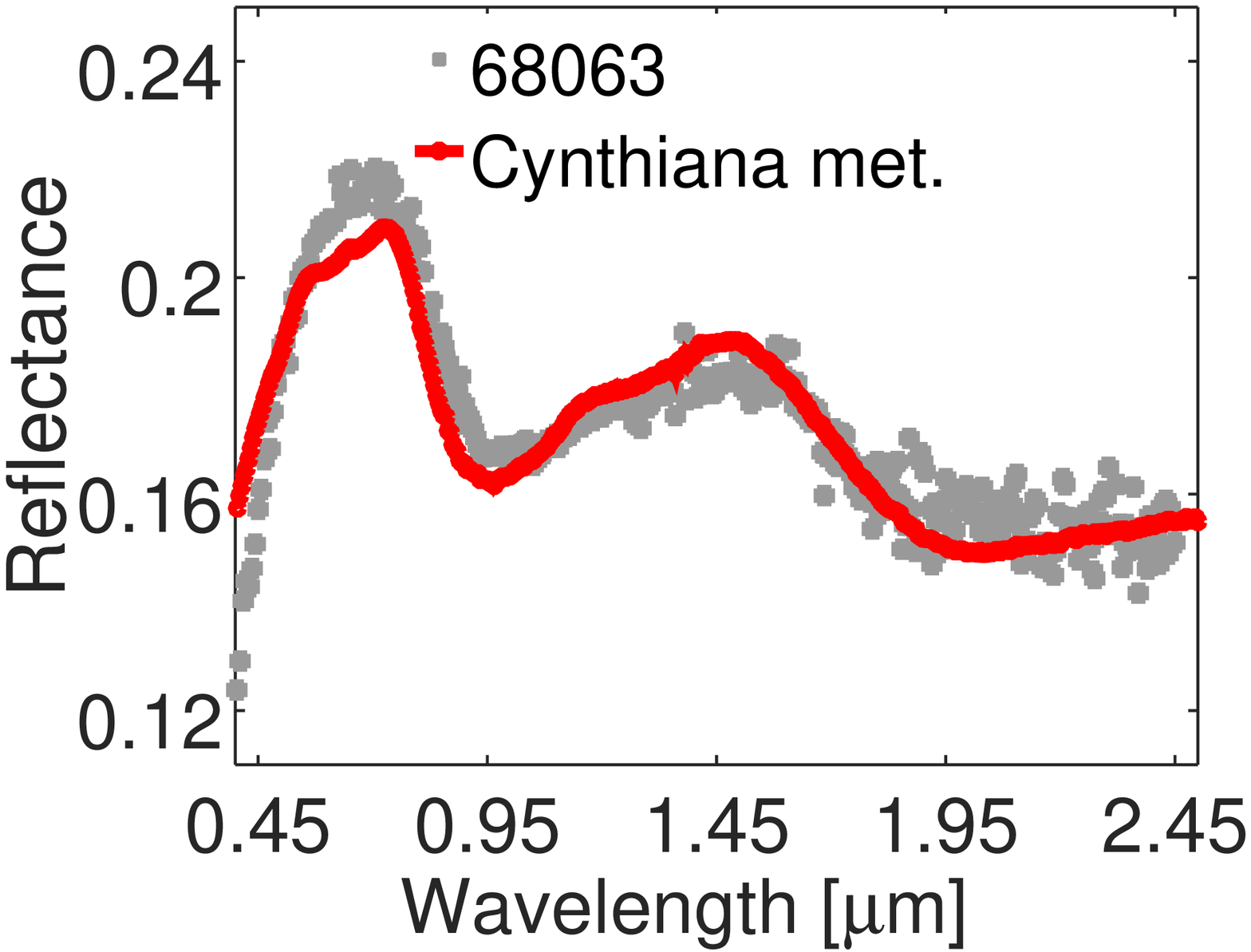}
\includegraphics[width=4.5cm]{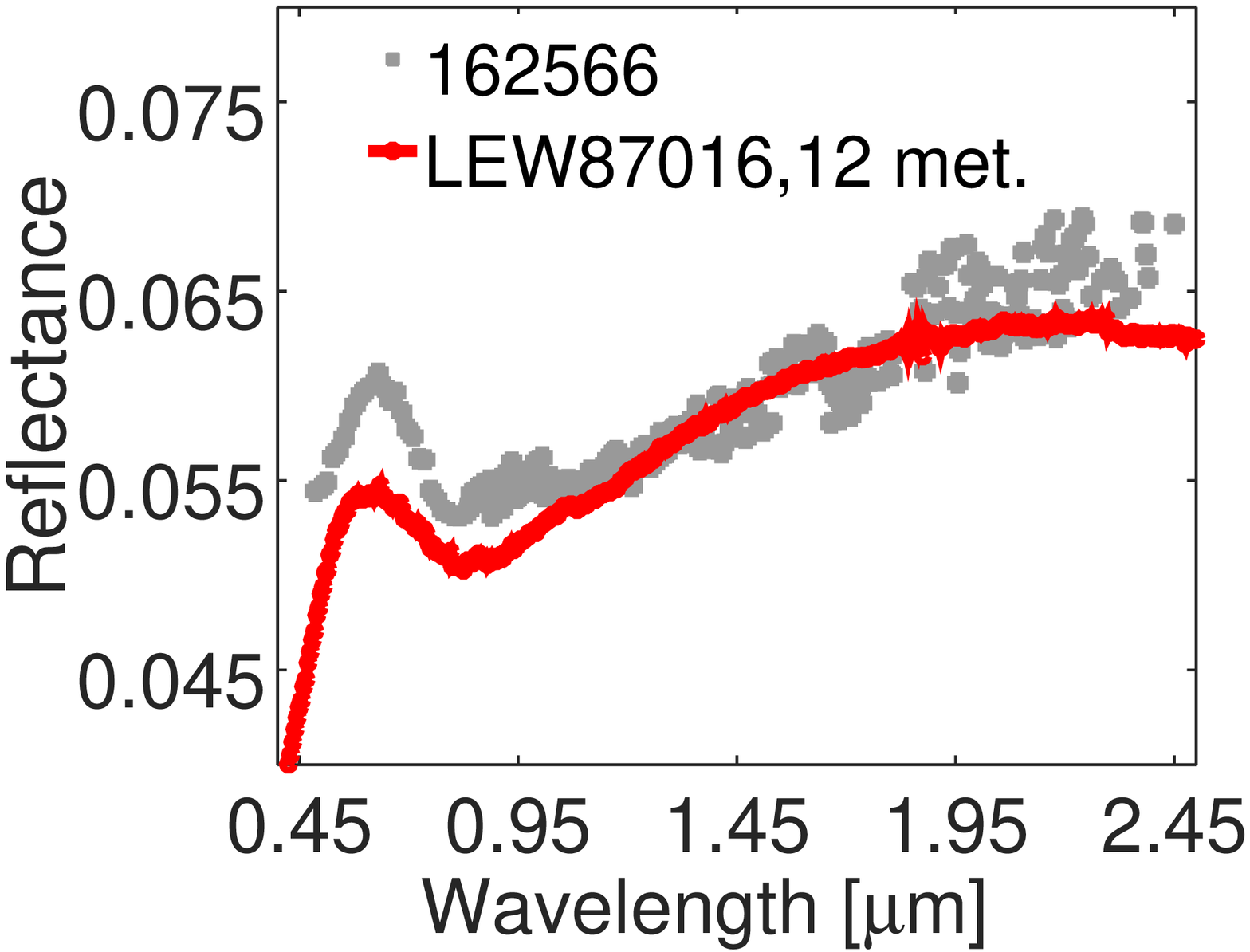}
\includegraphics[width=4.5cm]{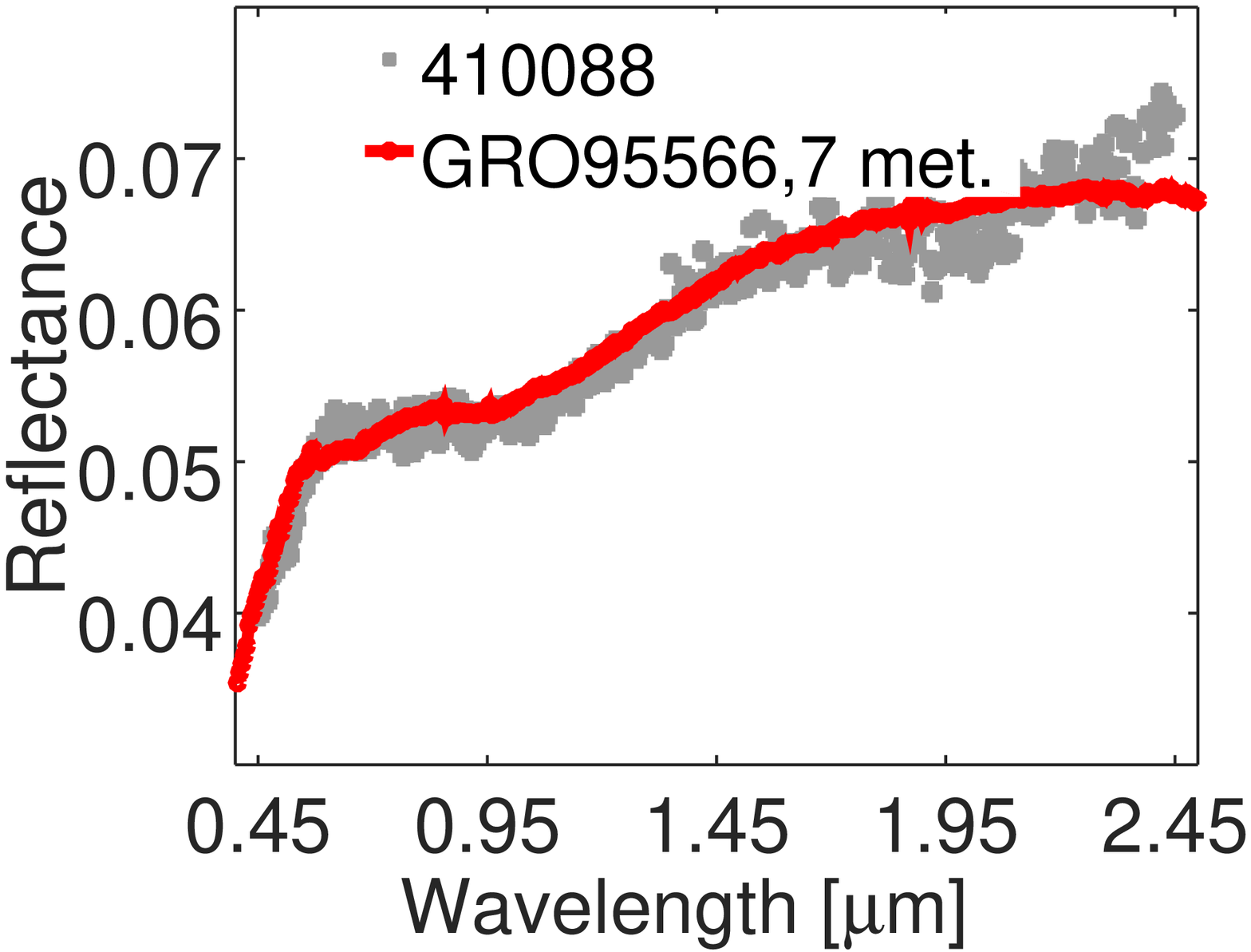}
\includegraphics[width=4.5cm]{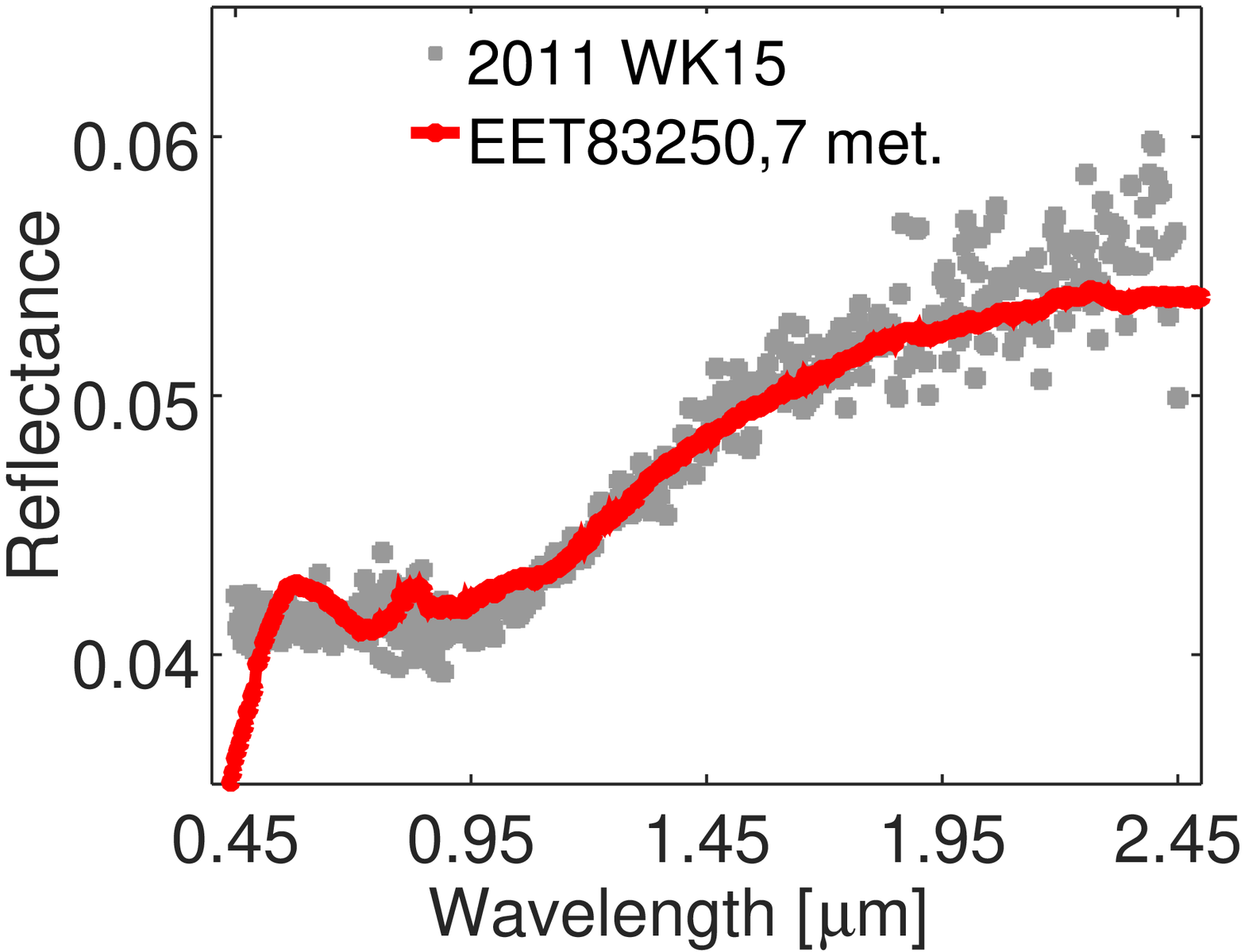}
\end{center}
\caption{ The best Relab meteorite spectral fit (shown in red) for the unusual asteroid spectral curves (grey) obtained from the concatenation of INT observations with SMASS data. The name of the asteroids and the name of the corresponding meteorite that matched are displayed in each figure. The asteroid spectrum was normalized to fit the meteorite albedo.}
\label{unusualspecmet}
\end{figure*}

\subsection{Uncommon spectra}

Several observed asteroids show unusual spectral features when compared with standard types. These spectra are shown in Fig.~\ref{unusualspec}. As we are not able to identify an observational error, we assume that these spectral curves are shaped by a peculiar composition, by other processes such as space-weathering, thermal emission, phase angle effects, or by the surface heterogeneity. To test if such spectra may represent a composition found in laboratory we compared them with the Relab database (Fig.~\ref{unusualspecmet}). However, more observations are required to validate these findings.

\emph{(68063) 2000 YJ66} is characterized by VNIR data that show the 1 $\mu$m band, but the blue slope in the wavelength interval 1.0 -- 2.5 $\mu$m is uncommon. The 2 $\mu$m  band associated to pyroxenes is not distinguishable (in the limit of the errorbars of the spectrum). The best meteorite spectral fit (Fig.~\ref{unusualspecmet}) of the combined VNIR spectrum is of a chip from L4 ordinary chondrite Cynthiana (sample ID OC-TXH-015-A).

The asteroid (68063) 2000 YJ66 is a large NEA, with an estimated diameter of 2.3 km. Its lightcurve shows a rotation period  of 2.11 hrs and a maximum amplitude of 0.14 mag. \citep{2015ApJ...814..117N}.

\emph{(162566) 2000 RJ34} has an uncommon spectrum in the optical region. Similar spectra  (Fig.~\ref{unusualspecmet}) are found in the Relab database and the best fit found for the optical part corresponds to LEW8701612  meteorite (sample ID MP-TXH-062). This sample is a C2 carbonaceous chondrite recovered at Lewis Cliff, Antarctica. In the NIR the curve fitting is poor, thus it prevents pointing to a compositional type. The albedo value $p_V = 0.071\pm0.015$ \citep{2011ApJ...743..156M} is in agreement with the carbonaceous like nature. Based on this value, its estimated size is 3.64 km. It is a slow rotator with a period of 50.5 $\pm$ 0.1 hrs. The lightcurve amplitude 0.78 $\pm$ 0.03 mag. indicates an elongated object \citep{2014MPBu...41..125W}.

\emph{(385186) 1994 AW1} is classified as a Cgh according to the INT spectrum. This observation was performed at a phase angle of 79$^{\circ}$. The NIR data from SMASS-MIT shows a feature at 2 $\mu$m (Fig.~\ref{unusualspec}). The complementary 1 $\mu$m feature for pyroxenes compositions is missing. A reconciliation of these two sets of data is to assume that the 2 $\mu$m feature of the NIR data is artificially generated and the increase of the spectral curve above 2 $\mu$m is rather a thermal tail.

This NEA is a binary object with a primary body about 600 m in diameter and a secondary being about half of the primary. The two asteroids orbit a common center of mass, at about 1.2 km apart \citep{2015DPS....4730808R}. \citet{2004Icar..170..259B} obtained a D-type like visible spectrum. If all of these observations are correct, the only explanation for the mismatch remains the surface heterogeneity, two bodies of very different composition, or both of these hypothesis. The next suitable geometry to observe this object will be in the summer of 2022. Because there are two observations in the visible region (i.e. ours and \cite{2004Icar..170..259B}) and one in NIR and there is a clear mismatch between them we did not analyze the combined spectrum relative to meteorites spectra.

\emph{(410088) 2007 EJ} has VNIR data similar to (385186) 1994 AW1. Its optical spectrum has an almost flat region covering 0.6-1.1 $\mu$m interval, and a drop in the blue region. This is characteristic for Cg taxonomic class. The NIR counterpart, retrieved from SMASS-MIT website matches the common interval with the INT data, but the slope over 1.1 - 1.8 $\mu$m is significantly redder compared to Cg- type. The increase in spectral curve above 2 $\mu$m can be attributed to thermal emission. If this thermal tail is ignored, several matches are found in the Relab database. The best one corresponds to "GRO95566,7" C2 carbonaceous chondrite (Fig.~\ref{unusualspecmet}). Several other spectral fits corresponding to samples from Murchison (e.g. sample ID: RS-CMP-045). The albedo value of $0.10\pm0.02$ determined by \citep{2015ApJ...814..117N} supports a Cg/Cgh type.

\emph{2011 WK15} shows different visible and near-infrared spectra. The visible data corresponds to a typical flat spectrum of a Cb-type asteroid. The NIR data corresponds to a D/T type asteroid, but the flat feature between 0.8 to 1.0 $\mu$m is unusual for these types. The spectral fit  (Fig.~\ref{unusualspecmet}) from the Relab database is of the C2 Carbonaceous Chondrite EET83250,7 (Sample ID: MP-TXH-040). However, this match is poor because the asteroid spectrum does not show the same features in the visible region as the meteorite. Rather, the visible part and the NIR one represent different compositions, or one of the observations has issues.

\section{Discussions and conclusions}

This article discusses the spectroscopic results of 76 NEAs having $T_J>3$. They were observed with INT/IDS instrument. The survey, performed as an EURONEAR collaboration, focused on large objects ($H\lesssim21$) and it complements other recent programs such as NEOShield-2 \citep{2018P&SS..157...82P} and MANOS \citep{2018DPS....5050801D, 2016AJ....152..163T}, which are oriented to observe small asteroids (i.e. smaller than 300 m). The wavelength interval covered is $\sim$0.4-1.0 $\mu$m. We enhanced our results for 33 of the targets using NIR data from SMASS-MIT website.

As of April 2019, the MPC lists the orbits of almost 20\,000 NEAs. The largest are (1036) Ganymede - 37.7 km,  (3552) Don Quixote - 19.0 km, (433) Eros - 16.8 km. We found information about spectral data or taxonomic classification for 1288 of these objects. The EARN database (August 2017 version), the SMASS-MIT website, and the M4AST database were used.  The analysis and the summary of taxonomic classification for about 1\,000 of NEAs were recently published by SMASS-MIT group \citep{2015aste.book..243B, 2019Icar..324...41B}.

We combined this information with the results of our survey to estimate the fraction of objects with taxonomic type determined versus the discovered ones as a function of their absolute magnitude. Fig.~\ref{neasize} shows that the NEAs with assigned spectral types are strongly dependent on their absolute magnitudes (and therefore their size). About half of the population with the estimated size about one kilometer (i.e. $H\sim17.5$ for $p_V=0.14$) is spectrally characterized in the literature, and only $\sim$4-10$\%$ for objects in the hundreds meter size range. Our survey contributed with about $6\%$ to the total number of NEAs with known asteroid spectra.

There are 27 NEAs from our sample categorized as PHAs. Their sizes are in the range of 160 - 2600 m. The estimated diameters for the largest of these objects are 2608 m for (90075) 2002 VU94, 2172 m -- (143992) 2004 AF, 1726 m -- (385186) 1994 AW1, 1717 m -- (159504) 2000 WO67. The frequent spectral types are Q (7 objects), Sq (4 objects), S (4 objects) and V (4 objects). Three of the five C-complex PHAs show a strong drop in the blue part of the spectrum corresponding to a Cg type. Even though this sample is small, the fraction of PHAs belonging to Q/S-complex (15/26) compared to those in C-Complex (5/26) is similar to the one obtained by \cite{2016AJ....151...11P} using a total of 262 objects (there are four asteroids in common with our dataset for which their classification is reported in Table.~\ref{physprop}) which included data from EARN database.

Among the asteroids which may endanger our planet, there is a set with a very similar orbit to Earth. These are the most suitable targets for spacecrafts. \citet{2015aste.book..855A} noted that one of the major goals for human spaceflight program is to identify an characterize the NEAs from the standpoint of human exploration and planetary defense. This will pave the way for Solar System exploration.

An important value for the feasibility of an asteroid to become a target for a space mission is the $\Delta V$ speed budget. This quantity is directly proportional to the amount of propellant required by a spacecraft with a given mass and propulsion system.  We determined the spectra of 31 low $\Delta V$ asteroids (i.e. $\Delta V~\leq~$7 km/s). Ten of these bodies are suitable for a sample-return mission ($\Delta V ~\leq~$6 km/s). The taxonomic distribution of the low  $\Delta V$ asteroids follows a similar trend as the one computed for all NEAs. About 15 of these objects belong to  Q/S-complex and 10 have carbonaceous chondrites like spectra. The noticeable ones are the end member spectral types  A -- (67367) 2000 LY27, (432655) 2010 XL69; B -- (275611) 1999 XX262, 2012 TM139, (249595) 1997 GH28; and V -- 2014 YB35, (348400)2005 JF21. 

In particular, the A-type asteroid (67367) 2000 LY27 requiring a $\Delta V$  of 5.131 km/s may represent a very good opportunity for space exploration of fragments coming from differentiated planetesimals in the early Solar System. Its visible spectrum suggests an olivine rich composition which may have formed in the mantle of a large body \citep{1998JGR...10313675S, 2007M&PS...42..155S}. These are very rare objects compared to the number predicted by the current models of Solar System formation, but they seem to be more abundant in the small asteroids population \citep{2018MNRAS.477.2786P}.

The asteroid (459872) 2014 EK24 has the lowest value of $\Delta V = 4.839$ km/s. It is one of the  NHATs\footnote{\url{https://cneos.jpl.nasa.gov/nhats/intro.html}} targets. We classified it as an S-type NEA, and we estimated its size in the order of 63 m. The NIR data retrieved from SMASS-MIT shows a red slope which points to a space-weathered object. The weak and shallow 2$\mu$m feature is an indication of high olivine content. Its fast rotation rate is in agreement with a monolithic body.

Another possible low $\Delta V $ monolithic body is (436724) 2011 UW158 ($\Delta V  = 5.105$ km/s). Its flat and featureless spectrum observed during this survey, and the fast rotation period are compatible with a metallic composition.

An overall conclusion of our survey is that the spectral types assigned to INT/IDS observed NEAs covered 20 classes out of the 24 defined by Bus-DeMeo taxonomy \citep{2009Icar..202..160D}. We determined 44 ($58\%$ out of the total sample) Q/S-complex, 16 ($21\%$) B/C-complex, 10 ($13\%$) V-types and 6($8\%$) miscellaneous types. This distribution shows a similar trend as the one reported by \citet{2004Icar..170..259B} and \citet{2015aste.book..243B}. It matches the compositional pattern shown by the inner main-belt population \citep{1982Sci...216.1405G, 2014Natur.505..629D}. We note that we did not identify D-, and T- types. These are very rare spectral types in the NEA population, and they are mostly found on cometary like orbits ($T_J\leq3$). Two of our spectra, corresponding to (276049) 2002 CE26  and (385186) 1994 AW1, shown the 0.7 $\mu$m feature which indicates the presence of hydrated minerals on their surface.

\citet{2019Icar..324...41B} and \citet{2018P&SS..157...82P} reported that the relative numbers of the taxonomic classes is not constant with the asteroids size. They found that that the distributions of NEAs with diameter below $\sim$ 250 m are different compared to those lager than this rough threshold. In particular, \citet{2018MNRAS.477.2786P} and \citet{2018MNRAS.476.4481B} found that the A-types and D-types are more abundant in the smaller NEOs. They used the 147 NEAs observed with  New Technology Telescope within the NEOShield-2 project \citep{2019Icar..324...41B}. Also, \citet{2019Icar..324...41B} shown that the Q/S-complex and V-types are dominating with about 80 $\%$ the mass distribution for NEOs < 200 m , and that B/C-complex asteroids are almost missing for this size range. However, they noted that  that  unaccounted bias  effects may contribute to the statistics of these bodies. Regarding the larger sizes, they concluded that the fractional distributions of major taxonomic classes (60 $\%$ S, 20 $\%$ C, 20 $\%$ other) "appear remarkably constant" over two orders of magnitudes in size $\sim$ 0.1 - 10 km. This finding is in agreement with our result.

Our NEAs observed sample outlines several connections between the spectral types of the NEAs reported in this paper and their orbital parameters. The median perihelion distance shows a pattern with respect to the most represented spectral classes. This can be quantified in terms of median value and median absolute deviation: the perihelion for the Q-types, $\bar{q_Q} = 0.797\pm0.244$ AU is the lowest, followed by the one of basaltic asteroids ($\bar{q_V} = 0.861\pm0.118$ AU). Commonly, the B/C-complex NEAs have a higher perihelion distance ($\bar{q_{Cc}} = 1.038 \pm 0.089$). This result is in agreement with the findings of \citet{2016Natur.530..303G} which used the albedo values. They reported that the deficit of low-albedo objects near the Sun is a result of a super-catastrophic breakup of a substantial fraction of asteroids when they achieve perihelion distances of a few tens of solar radii. They found that low-albedo asteroids break up more easily as a result of thermal effects and are more likely to be destroyed farther away from the Sun compared to other NEAs. Although, the NEAs in our observed sample can be categorized as large and they are more difficult to be completely disrupted, they fully follow the predictions of \citet{2016Natur.530..303G}.

A significant conclusion of our findings refers to the space-weathering process based on the relation between Q-types and S-complex asteroids. The NEAs classified in S-complex have the perihelia spread over large values, $\bar{q_{Sc}}=1.003\pm0.409$ AU.  However, the Q-types (which have similar albedos as S-complex, thus the same probability to be observed) have low perihelia and larger size. The large majority of asteroids with perihelion lower than 0.7 AU are Q-types. Moreover, the ratio of the Q-types is double compared to other results reported over the entire NEA's population (which include numerous small objects). These two evidences are complementary to the current models of  space-weathering effects \citep{2005Icar..173..132N, 2010Natur.463..331B, 2010Icar..209..510N, 2016Icar..268..340C, 2018Icar..304..162G}. To explain this observational result we took into account the model of \cite {2014Natur.508..233D} which show that the fresh regolith production by thermal fatigue fragmentation is an important process for the rejuvenation of NEA surfaces.

The B/C-complex asteroids come in two flavors with respect to their orbital inclination. Ten out of the 16 observed objects in this group have the inclination between 4.1$^\circ$ to 8.3$^\circ$ and are likely to originate in the inner main belt primitive families. Another group is formed by high-inclination (47-52 $^\circ$), large NEAs (2.0 -5.5 km estimated diameter), and with eccentricities $e\approx 0.5$. 

The visible part of the spectra of the B/C-complex asteroids fit with the standard spectral types. But the NIR part retrieved for some of the objects, from SMASS-MIT, is redder when compared with the typical C-complex spectra. This is at least the case of (385186) 1994 AW1, (410088) 2007 EJ, and 2011 WK15. Assuming no observational artifacts, this spectral may suggest heterogeneous compositions, weathering mechanism or a combination of both. Further observations are needed to confirm these findings.

Eight objects of our sample (representing more than 10$\%$) are identified as basaltic asteroids and classified as V-types. Their spectra are fitted by those of howardites, eucrites, diogenites meteorites. NIR spectral counterpart are available for five of them in the SMASS-MIT database. The computed band-parameters confirm their compositional similarity to HEDs achondrites.

\begin{acknowledgements}

The Isaac Newton Telescope is operated on the island of La Palma by the Isaac Newton Group of Telescopes in the Spanish Observatorio del Roque de los Muchachos of the Instituto de Astrof\'{\i}sica de Canarias. The authors acknowledges the Spanish and Dutch allocation committees for the following observing runs which allowed the data acquisition: N8/2014A, C97/2015A, N4/2015A, C26/2015B and N2/2015B. In addition, other data was acquired during some ING S/D nights (service time) and a few Spanish CAT service nights. We thank to the following ING students for helping us with some of the observations: C. Carlisle, R. Cornea, N. Gentile, F. Stevance, and V. Tudor.

MP acknowledges support from the AYA2015-67772-R (MINECO, Spain). JdL, JL, OV and MP acknowledge financial support from the project ProID20170112 (ACIISI/Gobierno de Canarias/EU/FEDER). The work of ILB was supported by a grant of the Romanian National Authority for Scientific Research - UEFISCDI, project number PN-III-P1-1.2-PCCDI-2017-0371. The work of MP, RMG and ILB  was supported by a grant of the Romanian National Authority for  Scientific  Research -- UEFISCDI,  project number PN-II-RU-TE-2014-4-2199. 
This research uses spectra acquired with the NASA Relab facility at Brown University. 
Part of the data used in this publication was obtained and made available by the the MIT-UH-IRTF Joint Campaign for NEO Reconnaissance. The IRTF is operated by the University of Hawaii under Cooperative Agreement no. NCC 5-538 with the National Aeronautics and Space Administration, Office of Space Science, Planetary Astronomy Program. The MIT component of this work is supported by NASA grant 09-NEOO009-0001, and by the National Science Foundation under Grants Nos. 0506716 and 0907766. Any opinions, findings, and conclusions or recommendations expressed in this material are those of the authors and do not necessarily reflect the views of MIT-UH-IRTF Joint Campaign. The authors thank to Dr. Beno\^{i}t Carry for the helpful comments that improved the paper.
\end{acknowledgements}

\bibliographystyle{aa}
\bibliography{NEAINT}

\clearpage
\newpage
\onecolumn

\begin{appendix}
\section{Summary of asteroid properties and circumstances of their observations}
{
\begin{longtable}{lccccccccl}
\caption{
\label{Circumstances}Log of observations: asteroid designation, apparent V magnitude ($m_V$), phase angle ($\Phi$), heliocentric distance ($r$), spectral slope -- $BR_{slope}(\%/0.1~\mu m)$, computed over 0.45 - 0.7 $\mu$m spectral interval, average seeing during the observation, start time (UT) of each observation, total exposure time, airmass and the solar analog (SA) used for data reduction are provided.}\\
\hline\hline
Asteroid & $m_V$ & $\Phi$ ($^\circ$) & $r~(UA)$ & $BR_{slope} $ & Seeing($"$) & $UT_{start}$ & $t_{exp}[s]$ & Airmass & SA \\
\hline
\endfirsthead
\caption{continued.}\\
\hline\hline
Asteroid & $m_V$ & $\Phi$ ($^\circ$) & $r~(UA)$ & $BR_{slope}$ & Seeing($"$) & $UT_{start}$ & $t_{exp}[s]$ & Airmass & SA \\
\hline
\endhead
\hline
\endfoot
1580      &15.2&63.0  &1.12463 &0.7  &1.1&2015-05-26T21:27&720&1.05&HD109967\\
4401      &17.1&29.7  &1.47579 &13.0 &1.2&2015-03-13T02:26&2700&1.10&HD100044\\
7889      &17.0&36.5  &1.52266 &25.6 &0.9&2015-05-28T04:00&1800&1.11&HD76617\\
16636     &17.5&5.4   &1.36513 &12.9 &0.8&2014-07-18T02:03&2500&1.51&HD195422\\
22099     &17.3&14.8  &1.35952 &10.4 &0.9&2015-03-11T04:18&2100&1.33&HD98562\\
25916     &17.5&32.0  &1.83638 &11.2 &0.9&2014-02-12T06:12&1200&1.38&HD127913\\
25916     &15.2&46.0  &1.36315 &18.3 &2.5&2014-05-13T04:54&2400&1.39&HD140235\\
39796     &16.8&56.1  &1.20896 &14.8 &1.0&2014-08-13T05:41&1200&1.15&HD206938\\
52750     &17.4&71.1  &1.05880 &12.8 &1.0&2015-05-11T21:58&3600&1.18&HD119550\\
66391     &16.9&71.9  &1.06593 &7.4  &1.1&2015-05-27T05:57&1800&1.06&HD142011\\
67367     &17.4&11.0  &1.55576 &25.9 &0.8&2015-05-24T02:27&1200&1.94&HD151928\\
68031     &17.1&34.0  &1.19811 &7.4  &0.9&2014-02-12T02:17&2400&1.11&HD120566\\
68063     &17.6&52.4  &1.27513 &11.1 &0.6&2014-08-21T04:24&2400&1.21&HD239928\\
86067     &18.2&31.4  &1.56370 &15.2 &0.7&2015-02-06T05:53&4000&1.01&HD109967\\
90075     &17.8&45.4  &1.41848 &6.9  &1.4&2014-07-28T05:16&800&1.05&HD9761\\
90075     &17.7&35.3  &1.61503 &2.8  &0.6&2014-08-21T03:15&4200&1.11&HD239928\\
90416     &16.3&41.6  &1.08886 &3.7  &0.7&2015-02-05T20:45&1200&1.08&HD76617\\
99248     &17.7&5.1   &1.78663 &14.6 &0.8&2014-12-30T02:01&3600&1.01&HD256516\\
103067    &16.2&46.7  &1.16413 &13.5 &0.7&2015-02-06T00:39&900&1.44&HD110485\\
112985    &17.5&32.6  &1.56425 &2.3  &1.1&2015-05-27T03:37&1800&1.45&HD182081\\
138127    &17.0&87.6  &0.96798 &5.8  &1.2&2015-03-12T22:45&1800&1.43&HD29714\\
138937    &16.3&7.0   &1.32307 &5.2  &2.0&2015-12-27T23:28&1200&1.11&HD218633\\
143992    &18.1&52.0  &1.24327 &16.2 &1.0&2015-02-06T21:11&2700&1.36&HYADE64\\
152564    &18.3&84.8  &1.01626 &17.0 &1.1&2015-05-26T23:29&3600&1.44&HD124019\\
152679    &16.7&40.6  &1.27486 &1.5  &0.9&2015-05-28T05:44&1500&1.23&HD182958\\
159504    &18.3&10.1  &1.76111 &2.3  &1.1&2015-05-27T01:28&3600&1.22&HD142011\\
162566    &15.9&59.0  &1.12403 &4.4  &0.9&2014-02-11T22:24&2520&1.22&HD284013\\
163696    &16.5&43.3  &1.23071 &17.4 &1.5&2015-12-25T21:43&1500&1.02&HD76617\\
190208    &18.2&6.3   &1.53949 &6.8  &0.6&2014-09-03T03:02&2700&1.25&HD217577\\
206378    &15.6&76.8  &1.02701 &8.1  &0.7&2015-08-31T05:57&800&1.16&HD218633\\
249595    &16.9&9.5   &1.35794 &1.8  &0.9&2014-02-12T01:05&3000&1.02&HD89084\\
267223    &18.0&47.1  &1.18551 &6.6  &1.5&2016-01-26T04:29&4800&1.09&HD257880\\
275611    &18.0&6.3   &1.50473 &-2.6 &0.6&2014-09-03T01:52&2700&1.17&HD216516\\
276049    &17.2&43.0  &1.29376 &5.1  &1.0&2014-08-13T03:52&1920&1.09&HD206938\\
276049    &14.1&24.3  &1.15085 &1.5  &0.6&2014-09-03T00:15&900&1.05&HD216516\\
285944    &13.2&29.0  &1.10719 &14.4 &0.6&2014-08-21T01:50&840&1.18&HD239928\\
294739    &17.9&24.2  &1.49875 &7.8  &0.8&2015-07-21T23:57&3000&1.04&HD157985\\
322763    &18.6&39.4  &1.42111 &14.1 &1.2&2015-03-12T23:19&4800&1.15&HD75488\\
333578    &17.2&66.8  &1.04784 &14.5 &0.6&2014-08-21T02:41&2100&1.18&HD239928\\
337866    &15.6&21.6  &1.09789 &11.4 &1.5&2015-12-25T00:35&900&1.39&HD76617\\
348400    &18.2&24.2  &1.52664 &17.0 &0.8&2015-05-24T03:24&2400&1.13&HD170717\\
348400    &14.8&32.7  &1.14018 &18.5 &0.8&2015-07-21T22:05&540&1.17&HD131715\\
357439    &16.3&50.7  &1.03929 &14.9 &0.7&2015-02-06T03:31&2400&1.29&HD110485\\
380981    &17.5&66.0  &1.07347 &10.8 &0.8&2015-05-24T04:12&2400&1.06&HD231683\\
381677    &18.0&60.4  &1.09503 &8.4  &0.9&2014-02-12T05:14&3600&1.17&HD120566\\
385186    &17.9&79.1  &1.02465 &2.0  &0.7&2015-08-30T22:04&2700&1.30&HD198273\\
389694    &16.7&23.2  &1.24496 &16.0 &1.0&2015-09-28T03:20&2400&1.05&SA93-101\\
391033    &17.9&32.3  &1.21338 &17.4 &0.8&2014-07-18T00:21&3600&1.28&HD195422\\
398188    &16.5&57.4  &1.06592 &11.0 &1.4&2014-07-28T05:48&1200&1.03&HD9761\\
399307    &17.6&15.4  &1.29125 &5.4  &0.6&2014-09-03T02:17&1800&1.10&HD217577\\
410088    &17.3&42.2  &1.17319 &10.2 &0.7&2015-02-05T23:45&2700&1.18&HD76617\\
410195    &17.8&36.9  &1.21250 &8.4  &0.8&2014-12-30T01:38&2400&1.33&HD19061\\
410627    &16.8&18.4  &1.10103 &19.0 &0.6&2014-09-03T02:42&1800&1.73&HD221227\\
411280    &17.6&9.6   &1.25266 &3.4  &0.9&2014-02-11T19:55&2400&1.15&HD73708\\
416151    &17.4&43.9  &1.05193 &5.9  &0.7&2015-02-06T04:36&3500&1.06&HD109967\\
416224    &18.3&31.7  &1.37227 &9.4  &0.9&2015-05-27T23:23&3300&1.03&HD76617\\
419022    &18.0&28.9  &1.27138 &9.4  &0.9&2014-02-12T04:41&3600&1.04&HD120566\\
423747    &17.4&21.9  &1.22160 &16.8 &1.2&2015-03-13T04:24&3000&1.54&HD118928\\
425450    &18.1&39.2  &1.16370 &10.7 &1.1&2015-05-26T22:11&3600&1.19&HD142011\\
429584    &17.9&49.2  &1.07499 &15.9 &0.7&2015-02-06T06:08&2700&1.49&HD110485\\
430544    &16.8&41.8  &1.10395 &7.4  &0.9&2015-03-11T05:03&1400&1.58&HD121867\\
432655    &18.6&22.9  &1.26368 &18.6 &0.9&2015-05-28T02:18&3600&1.12&HD76617\\
436724    &17.6&44.6  &1.10774 &1.9  &0.7&2015-08-31T01:22&2700&1.06&SA110-361\\
436775    &15.8&26.6  &1.28305 &4.1  &1.0&2015-05-11T22:57&1600&1.08&HD76617\\
438429    &18.2&48.8  &1.16689 &3.9  &0.7&2015-08-31T04:40&3000&1.01&SA93-101\\
443103    &17.1&103.8 &0.97831 &9.3  &0.6&2014-09-02T21:00&2400&1.33&HD216516\\
450263    &18.1&40.1  &1.18932 &7.9  &1.5&2015-12-26T06:58&3600&1.04&HD89084\\
455554    &17.5&38.5  &1.21098 &7.8  &1.5&2016-01-25T21:45&1800&1.32&HD76617\\
459872    &18.0&18.5  &1.04004 &13.8 &1.2&2015-03-13T01:37&2700&1.25&HD91658\\
463380    &17.9&15.3  &1.27523 &14.0 &1.5&2016-01-26T01:40&2400&1.13&HD79078\\
2007 ED125&16.7&37.5  &1.03070 &8.7  &0.9&2015-03-11T06:37&1200&1.55&HD118928\\
2008 JY30 &18.6&62.8  &1.08418 &8.3  &2.0&2015-12-27T20:24&4800&1.15&HD218633\\
2011 WK15 &18.3&16.6  &1.24728 &-0.1 &0.8&2014-12-30T05:44&3600&1.17&HD73708\\
2012 TM139&17.6&25.7  &1.18259 &0.2  &0.7&2015-08-30T23:01&2250&1.10&SA110-361\\
2013 AV60 &17.3&48.5  &1.11150 &13.8 &2.0&2015-12-27T22:45&2400&1.20&HD254642\\
2014 YB35 &18.2&43.6  &1.16213 &28.6 &0.7&2015-02-06T00:01&1000&1.33&HD109967\\
2014 YM9  &18.4&69.8  &1.04191 &7.9  &0.9&2015-03-11T06:09&1200&1.34&HD121867\\
2015 HA1  &17.5&44.6  &1.06639 &12.9 &0.8&2015-05-23T22:50&3600&1.37&HD110747\\
2015 HP43 &17.6&6.7   &1.14592 &14.5 &0.8&2015-05-24T01:16&3000&1.43&HD137338\\
2015 SV2  &17.7&12.8  &1.12305 &4.0  &2.0&2015-12-28T01:40&2400&1.07&HD259516\\
\hline
\end{longtable}
}

{
\begin{longtable}{lccccccclccc}

\caption{\label{physprop} Summary of the properties of the NEAs discussed in this article. The \emph{Tax.} column shows the taxonomic type determined in this work. It is compared with the previous taxonomic type (\emph{Tax.Prev.}) and albedo ($p_V$) reported in the EARN database (when available). The absolute magnitude (\emph{H}, the orbit type (\emph{Orbit}) -- Amor (AM), Apollo (AP), Aten (AT), and the \emph{MOID} are retrieved from NEODyS. The '*' symbol on the \emph{Orbit} column marks the PHA. The equivalent diameters (\emph{D}) were computed as described in the Methods section. The rotation period (\emph{Psyn}) and the maximum lightcurve amplitude (\emph{A}) we retrieved from LCDB databse. The $\Delta V$ budget and the Tisserand parameter ($T_J$) were retrieved from the websites of Lance A. M. Benner. The information from EARN databse was obtained on August 2017, while for all the others we used the version of September 2018. The references corresponding to \emph{Tax.Prev. column are: 
(1) - \cite{2005Icar..175..175R}, 
(2) - \cite{2001PhDT.......121W}, 
(3) - \cite{2014Icar..227..112D}, 
(4) - \cite{2004Icar..170..259B}, 
(5) - \cite{2014Icar..228..217T}, 
(6) - \cite{2011AJ....141...32Y}, 
(7) - \cite{2015A&A...581A...3B}, 
(8) - \cite{2002Icar..158..146B}},
(9) - \cite{2016AJ....151...11P}
}\\
\hline\hline       
Designation &Tax.&Tax.Prev.&$H$&$p_V$&$D[m]$&$Psyn[h]$& $Amp[mag]$ & Orbit & $\Delta V[km/s]$ & $T_J$  & $MOID[au]$ \\ 
\hline		  
\endfirsthead
\caption{continued.}\\
\hline\hline
Designation &Tax.&Tax.Prev.&$H$&pv&$D[m]$&$Psyn[h]$& $A[mag]$ & Orbit & $\Delta V[km/s]$ & $T_J$  & $MOID[au]$ \\ 
\hline
\endhead
\hline
\endfoot
1580       & B  & C$^{(1)}$    & 14.7 &0.08&5499&6.132&0.70&AM&16.06&3.065&0.13566\\
4401       & Sr & -          & 16.0 & - &1626&6.670&0.64&AM&9.49&3.056&0.33021\\
7889       & V  & V$^{(2)}$    & 15.2 &0.52&1681&2.741&0.39&AP&12.92&4.865&0.15506\\
16636      & Sq & Scomp$^{(3)}$& 18.7 & - &491&19.010&0.41&AM&7.04&3.421&0.24073\\
22099      & Sq & S$^{(4)}$    & 17.9 &0.29&649&6.334&0.41&AP&6.32&5.585&0.1676\\
25916      & Sq & Sq$^{(5)}$   & 13.6 &0.21&5593&4.190&0.37&AM&8.22&3.203&0.28042\\
39796      & Q  & -          & 15.7 &0.20&2153&223.500&0.92&AM&7.39&3.445&0.18991\\
52750      & V  & V$^{(5)}$    & 16.4 &0.39&1113&26.430&0.24&AP&7.60&4.520&0.16143\\
66391      & Q  & S, Sa$^{(3,9)}$ & 16.5 & - &1398&2.765&0.12&AT*&21.31&8.500&0.01337\\
67367      & A  & -          & 17.0 & - &1211& - & - &AM*&5.14&4.943&0.04596\\
68031      & Q  & -          & 18.1 & - &669&2.550&0.06&AM&7.37&4.765&0.23079\\
68063      & Q  & -          & 15.5 &0.21&2304&2.110&0.14&AM&7.09&3.415&0.27422\\
86067      & Sa & S$^{(5)}$    & 16.4 &0.31&1245&9.196&0.55&AM&10.77&3.827&0.36552\\
90075      & Q  & -          & 15.3 &0.20&2608&7.879&0.65&AP*&6.66&3.472&0.03056\\
90416      & Cg & -          & 18.6 & - &1009&43.580&0.13&AP*&6.27&4.054&0.00132\\
99248      & S  & S,T$^{(6)}$   & 16.4 &0.42&1076&19.700&0.30&AP*&6.42&3.802&0.04695\\
103067     & Q  & S,Sr$^{(9,3)}$ & 16.8 &0.18&1387&9.849&0.49&AP*&12.47&3.580&0.04094\\
112985     & C  & -          & 15.7 & - &4306&4.787&0.24&AM&14.53&3.119&0.41253\\
138127     & Q  & Q$^{(2,9)}$    & 17.1 & - &1061&2.586&0.26&AT*&16.51&8.399&0.02181\\
138937     & Q  & X$^{(6)}$    & 17.5 &0.21&917& - & - &AP&11.28&3.298&0.06052\\
143992     & L  & -          & 16.1 & - &2075& - & - &AP*&9.33&3.552&0.03258\\
152564     & Sv & -          & 19.8 &0.17&354&3.276&0.13&AP&8.38&4.573&0.12034\\
152679     & Xc & Cb$^{(4)}$   & 16.6 &0.02&4741&125.000&1.35&AM&6.12&3.403&0.0598\\
159504     & Xk & -          & 17.0 & - &1717&7.840&0.20&AP*&6.92&3.198&0.04844\\
162566     & Cg & X$^{(6)}$    & 15.7 &0.07&3639&50.670&0.78&AM&7.59&3.105&0.17779\\
163696     & V  & -          & 16.5 & - &1107&62.400&1.64&AP&10.42&4.129&0.05278\\
190208     & Cg & -          & 18.1 & - &1270&182.000&0.25&AM&5.99&3.628&0.09711\\
206378     & Sq & -          & 18.7 & - &491&37.500&0.43&AM*&5.51&3.956&0.0445\\
249595     & B  & -          & 17.7 & - &1106&19.696&1.20&AM&6.80&3.740&0.30668\\
267223     & Q  & -          & 18.1 & - &669&4.110&0.38&AP&12.19&3.326&0.10641\\
275611     & B  & -          & 18.0 &0.03&2111&15.112&0.16&AM&6.33&4.450&0.31114\\
276049     & Cgh,Cg & -          & 16.8 & - &2311&3.293&0.07&AP&14.09&3.065&0.09712\\
285944     & V  & -          & 16.5 &0.32&1177&2.246&0.11&AM&16.17&3.045&0.08929\\
294739     & Sq & -          & 17.1 & - &1025&3.054&0.49&AP*&11.57&4.133&0.04313\\
322763     & S  & -          & 17.1 &0.20&1130& - & - &AP&8.30&3.558&0.10127\\
333578     & S  & -          & 20.2 & - &264&5.737&0.71&AP*&5.78&4.303&0.01307\\
337866     & Q  & Sk$^{(4)}$   & 18.8 & - &485&8.955&0.17&AM&6.01&3.699&0.06653\\
348400     & V  & -          & 17.3 & - &766&2.415&0.14&AM*&6.60&3.423&0.04724\\
357439     & V  & V$^{(7)}$    & 19.4 & - &291&2.621&0.17&AP*&8.46&4.364&0.00743\\
380981     & Q  & -          & 18.7 & - &508&4.920&0.18&AM*&5.85&4.056&0.03324\\
381677     & Q  & -          & 18.3 &0.30&531& - & - &AM&5.58&3.867&0.09924\\
385186     & Cgh& Sa$^{(8,9)}$   & 17.6 & - &1574&2.519&0.17&AM*&10.24&5.547&0.01954\\
389694     & S  & -          & 18.1 &0.25&638&5.240&0.08&AP*&8.07&4.264&0.01937\\
391033     & S  & -          & 19.1 & - &438&850.000&1.00&AM&6.56&3.605&0.19363\\
398188     & Sq & -          & 19.5 &0.15&432&21.990&1.12&AT*&8.37&6.785&0.02235\\
399307     & X  & -          & 18.9 & - &1017&3.481&0.09&AM&7.18&3.519&0.26562\\
410088     & Cg & -          & 18.1 &0.10&1008&4.781&0.16&AP&6.90&3.147&0.05843\\
410195     & Q  & -          & 18.4 & - &583&48.050&1.08&AM&6.80&3.441&0.23077\\
410627     & V  & -          & 20.7 & - &160& - & - &AP*&7.34&4.832&0.01316\\
411280     & Xn & -          & 19.3 & - &551&7.790&0.20&AM&6.30&3.723&0.19864\\
416151     & C  & C/X$^{(5)}$  & 20.7 & - &431&12.191&0.72&AP*&5.95&5.556&0.04989\\
416224     & Sv & -          & 18.0 & - &601&7.666&1.02&AM&7.84&3.924&0.31827\\
419022     & Sv & -          & 18.5 & - &477&4.840&0.75&AM&15.96&4.284&0.28775\\
423747     & Sv & -          & 18.9 & - &397&3.200&0.10&AM&10.38&3.994&0.22639\\
425450     & Sq & -          & 19.6 & - &324&3.520&0.27&AM&6.33&3.909&0.19459\\
429584     & S  & -          & 19.9 & - &303&43.500&0.65&AP*&7.47&3.071&0.00297\\
430544     & Q  & -          & 18.6 & - &532&2.633&0.41&AP*&9.27&3.132&0.01078\\
432655     & A  & -          & 19.7 & - &349& - & - &AM&6.79&4.435&0.27825\\
436724     & C  & -          & 19.9 & - &360&0.611&2.05&AM*&5.11&4.241&0.002\\
436775     & O  & -          & 16.5 & - &1144&5.687&0.41&AP&12.84&3.250&0.22125\\
438429     & Xk & -          & 18.9 & - &716&3.686&0.22&AM&6.39&3.610&0.16149\\
443103     & Cg & -          & 18.0 & - &1330&135.000&1.10&AP*&9.30&5.841&0.01297\\
450263     & Sq & -          & 18.8 &0.34&396& - & - &AP*&7.24&4.564&0.04727\\
455554     & Xe & -          & 18.0 & - &905& - & - &AP&8.07&3.103&0.12145\\
459872     & S  & -          & 23.3 & - &63&0.098&1.26&AP&4.84&6.041&0.03173\\
463380     & Sr & -          & 19.1 & - &390&425.000&0.49&AM&10.28&3.482&0.30561\\
2007 ED125 & Q  & -          & 21.0 & - &176&5.620&0.55&AP*&6.74&3.689&0.00911\\
2008 JY30  & Q  & -          & 18.9 & - &463& - & - &AP&8.74&3.056&0.17652\\
2011 WK15  & C  & -          & 19.7 & - &682&3.541&0.37&AM&6.05&3.695&0.1126\\
2012 TM139 & B  & -          & 19.7 & - &440&2.680&0.18&AM&6.52&3.599&0.17879\\
2013 AV60  & Q  & -          & 18.6 & - &532&3.200&0.22&AP&8.54&3.370&0.10843\\
2014 YB35  & V  & -          & 19.0 & - &350&3.277&0.16&AP*&6.37&3.800&0.02174\\
2014 YM9   & S  & -          & 19.3 & - &399&7.300&0.04&AM&7.52&3.750&0.10011\\
2015 HA1   & S  & -          & 21.2 & - &166&47.200&0.39&AT&8.91&6.192&0.05539\\
2015 HP43  & Sr & -          & 21.1 & - &155&5.770&0.11&AM&7.76&3.412&0.13228\\
2015 SV2   & Xc & -          & 20.8 & - &256&43.910&0.80&AM&6.63&3.227&0.09575\\
\hline
\end{longtable}
}

\twocolumn
\section{The equivalent designation of asteroids (as of March 31, 2019)}

{\begin{itemize}
\item[$\star$] 1580, 1950 KA, Betulia
\item[$\star$] 4401, 1985 TB, Aditi
\item[$\star$] 7889, 1994 LX
\item[$\star$] 16636, 1993 QP, 2000 NR3
\item[$\star$] 22099, 2000 EX106, 1994 BX4
\item[$\star$] 25916, 2001 CP44, 1973 GM
\item[$\star$] 39796, 1997 TD
\item[$\star$] 52750, 1998 KK17
\item[$\star$] 66391, 1999 KW4
\item[$\star$] 67367, 2000 LY27
\item[$\star$] 68031, 2000 YK29
\item[$\star$] 68063, 2000 YJ66, 1964 VG
\item[$\star$] 86067, 1999 RM28
\item[$\star$] 90075, 2002 VU94
\item[$\star$] 90416, 2003 YK118
\item[$\star$] 99248, 2001 KY66
\item[$\star$] 103067, 1999 XA143
\item[$\star$] 112985, 2002 RS28
\item[$\star$] 138127, 2000 EE14
\item[$\star$] 138937, 2001 BK16
\item[$\star$] 143992, 2004 AF
\item[$\star$] 152564, 1992 HF
\item[$\star$] 152679, 1998 KU2, 1971 UB
\item[$\star$] 159504, 2000 WO67
\item[$\star$] 162566, 2000 RJ34
\item[$\star$] 163696, 2003 EB50
\item[$\star$] 190208, 2006 AQ
\item[$\star$] 206378, 2003 RB
\item[$\star$] 249595, 1997 GH28
\item[$\star$] 267223, 2001 DQ8
\item[$\star$] 275611, 1999 XX262
\item[$\star$] 276049, 2002 CE26
\item[$\star$] 285944, 2001 RZ11
\item[$\star$] 294739, 2008 CM
\item[$\star$] 322763, 2001 FA7
\item[$\star$] 333578, 2006 KM103
\item[$\star$] 337866, 2001 WL15
\item[$\star$] 348400, 2005 JF21, 2005 JU67
\item[$\star$] 357439, 2004 BL86
\item[$\star$] 380981, 2006 SU131
\item[$\star$] 381677, 2009 BJ81
\item[$\star$] 385186, 1994 AW1
\item[$\star$] 389694, 2011 QD48
\item[$\star$] 391033, 2005 TR15
\item[$\star$] 398188, 2010 LE15, Agni
\item[$\star$] 399307, 1991 RJ2
\item[$\star$] 410088, 2007 EJ
\item[$\star$] 410195, 2007 RT147
\item[$\star$] 410627, 2008 RG1
\item[$\star$] 411280, 2010 SL13
\item[$\star$] 416151, 2002 RQ25
\item[$\star$] 416224, 2002 XM90
\item[$\star$] 419022, 2009 QF31
\item[$\star$] 423747, 2006 CX
\item[$\star$] 425450, 2010 EV45, 2005 ME55
\item[$\star$] 429584, 2011 EU29
\item[$\star$] 430544, 2002 GM2
\item[$\star$] 432655, 2010 XL69
\item[$\star$] 436724, 2011 UW158
\item[$\star$] 436775, 2012 LC1
\item[$\star$] 438429, 2006 WN1
\item[$\star$] 443103, 2013 WT67
\item[$\star$] 450263, 2003 WD158
\item[$\star$] 455554, 2004 MQ1
\item[$\star$] 459872, 2014 EK24
\item[$\star$] 463380, 2013 BY45, 2010 GD7
\item[$\star$] 2007 ED125
\item[$\star$] 2008 JY30, 2015 HO9
\item[$\star$] 2011 WK15
\item[$\star$] 523667, 2012 TM139
\item[$\star$] 2013 AV60
\item[$\star$] 523775, 2014 YB35
\item[$\star$] 2014 YM9
\item[$\star$] 2015 HA1
\item[$\star$] 2015 HP43
\item[$\star$] 2015 SV2
\end{itemize}
}

\clearpage
\section{Comparison between spectra obtained at different dates for the same object}
\begin{figure}
\begin{center}
\includegraphics[width=8cm]{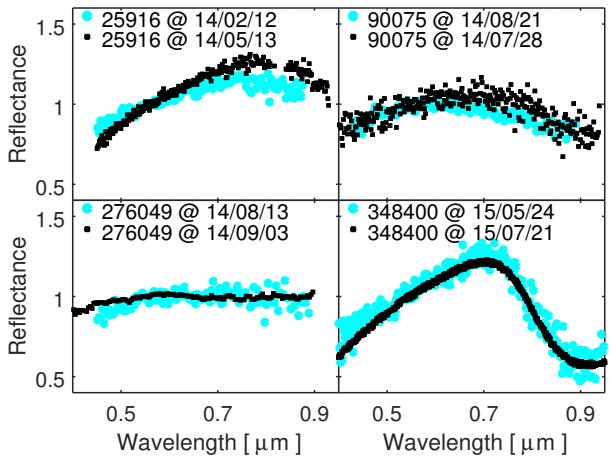}
\end{center}
\caption{The spectrum presented in Fig.~\ref{VisSpectra} is shown with black, while the second spectrum is shown with cyan. The legend of the plot show the name of the object and the observation date in the format yy/mm/nn (yy-year, mm-month, nn-night of observation)}
\label{doublespec}
\end{figure}

\end{appendix}
\end{document}